\documentclass{emulateapj}

\slugcomment{To appear in ApJ, vol. 699 (June 20, 2009, issue)}

\shorttitle{H$\alpha$ Galaxy Survey. VIII. Close companions, interactions, and the definition of starbursts}
\shortauthors{J. H. Knapen \& P. A. James}

\usepackage{graphicx}

\newcommand{\Ha}{H$\alpha$}
\newcommand{\ha}{H$\alpha$}
\newcommand{\hi}{H{\sc i}}

\begin{document}


\title{The H$\alpha$ Galaxy Survey. VIII. Close companions and interactions, and the definition of starbursts}


\author{Johan H. Knapen}\affil{Instituto de Astrof\'\i sica de
Canarias, E-38200 La Laguna, Spain; jhk@iac.es}

\author{Philip A. James}\affil{Astrophysics Research
	  Institute, Liverpool John Moores University, Twelve Quays
	  House, Egerton Wharf, Birkenhead CH41 1LD, UK}



\begin{abstract}

We consider the massive star formation properties, radial profiles, and atomic gas masses of those galaxies in our \ha\ Galaxy Survey that have close companion galaxies, in comparison with a matched control sample of galaxies without companions. Our analysis is based on \ha\ and continuum images of 327 disk galaxies that form a representative sample of the local Universe. We find that the presence of a close companion raises the star formation rate by a factor of just under two, while increasing hardly at all the equivalent width of the H$\alpha$ emission. This means that although statistically galaxies with close companions form stars at a higher rate, they do this over extended periods of time, and not as bursts. We find no significant increase in the central concentration of the star formation as a result of the presence of a close companion. 

The fraction of truly interacting or merging galaxies is very small in the local Universe, at around 2\%, and possibly 4\% of bright galaxies. Most of these interacting galaxies currently have unremarkable star formation properties.

We also study the properties of the galaxies in the Survey with the most extreme values for star formation indicators such as rate, equivalent width, star formation rate per area, and gas depletion timescale. We find that each of these indicators favors a different subset of galaxies, and use this information to discuss critically the possible definitions of the term starburst to describe galaxies with enhanced star formation activity. We conclude that no one starburst definition can be devised which is objective and generally discriminant. Unless one restricts the use of the term ``starburst'' to a very small number of galaxies indeed, the term will continue to be used for a heterogeneous and wide-ranging collection of objects with no physical basis for their classification as starburst.

Our overall conclusions are that (1) whereas the rare interacting and merging galaxies may have enhanced star formation, and vice versa, those galaxies with the highest star formation are often interacting or merging, the influence of the presence of a close companion on the star formation in galaxies is in general very small, and long-lasting; and (2) the term ``starburst'' cannot be easily defined objectively and generally in physical terms.

\end{abstract}




\section{Introduction}\label{intro}

Ample anecdotal evidence exists for a causal connection between galaxy interactions and enhanced star formation (SF), the latter often referred to by the generic term starburst. Practically all of the most luminous infrared sources, thought to be powered mainly by extreme starbursts (e.g., Genzel et al. 1998), occur in galaxies which show clearly disturbed morphologies, and which are presumably interacting or merging. This is in particular the case for the ultra-luminous infrared galaxies (ULIRGs, with $L_{\rm IR}=L[8-1000\,\mu{\rm m}]>10^{12}\,L_\odot$; see, e.g., Joseph \& Wright 1985; Armus, Heckman, \& Miley 1987; Sanders et al. 1988; Clements et al. 1996; Sanders \& Mirabel 1996 ) and the luminous infrared galaxies (LIRGs, $10^{11}<L_{\rm IR}<10^{12}\,L_\odot$; see, e.g., Scoville et al. 2000; Arribas et al. 2004; Sanders \& Ishida 2004; Alonso-Herrero et al. 2006). But these kinds of galaxies are indeed extreme, and very rare, at least at low redshifts. In addition, the much enhanced SFRs seen in these objects may be the result of mergers, rather than pre-merger interactions (or interactions not leading to mergers at all). 

In order to assess the more general impact of interactions and mergers on SF in galaxies, one must thus look beyond the (U)LIRGs, and there the picture is much less clear. In a classical paper, Larson \& Tinsley (1978) constructed stellar population models to explain the significantly larger scatter in the ($U-B, B-V$) diagram for peculiar galaxies from the Arp catalogue (Arp 1966) as compared to normal galaxies from the Hubble atlas. They inferred that this behavior can be explained by large variations in the SF rate (SFR) on short timescales, called bursts of SF, and noted that such bursts can in particular be identified with galaxies undergoing violent dynamical phenomena---interacting galaxies. 

Although many studies corroborated the basic conclusion from the important work of Larson \& Tinsley (1978) that galaxy interactions cause an increase in the SFR (see, e.g., Bushouse 1987; Kennicutt et al. 1987; and Keel 1991 and Struck 1999, 2007 for reviews), there are dissonant views. For instance, Bergvall, Laurikainen, \& Aalto (2003) find that the optical colors  are very similar in matched samples of merging/interacting and non-interacting galaxies, and Brosch, Almoznino, \& Heller (2004) find that the SFR is not significantly enhanced in interacting dwarf galaxies. 

From simulations we know that although galaxy interactions can lead to enhanced SFRs, they certainly do not do so in all circumstances, and the SFR is often only enhanced by a factor of two, or a few (e.g., Mihos \& Hernquist 1996; Kapferer et al. 2005; Di Matteo et al. 2007; Cox et al. 2008). Di Matteo et al. (2008) find from many sets of simulations of interacting and merging galaxies obtained with different numerical codes that in by far most cases the SFR is only enhanced by factors of a few, and that the duration of this enhancement is limited to a few hundred million years. The limited increase in SFR had been reported as early as 1987 by Bushouse (who found that the overall SFR as derived from \ha\ was enhanced by a factor of 2.5 in interacting galaxies, a factor which rose to six when he considered the SFR as derived from IR observations), and seems to agree with new observational results by, among others, Smith et al. (2007), Woods \& Geller (2007), Li et al. (2008) and Jogee et al. (2008). Below, we will present further affirmative evidence.

In the present paper, we focus on two related questions, namely what the role of interactions and close companions is in determining the SFR of local galaxies, and how we can define the term ``starburst". We use the sample of galaxies from the \ha\ Galaxy Survey (H$\alpha$GS; James et al. 2004, hereafter paper~I) for this purpose, a statistically meaningful sample of 327 local galaxies, many of them dwarfs and/or irregulars. From Paper~I, we have at our disposal the integrated SFR and \ha\ equivalent width (EW) measured across each galaxy, as well as the radial variations in these quantities. 
Because merging and interacting galaxies are very rare in the local Universe, we widen the scope somewhat and consider also galaxies that have a close companion. 

Our sample of 327 galaxies is small compared to samples derived from surveys such as the Sloan Digital Sky Survey (SDSS), and we need to discuss the pros and cons of using the H$\alpha$GS for this. To illustrate the power of the SDSS, we mention two examples of recent work in this area. First, Li et al. (2008) use a sample of $10^5$ galaxies selected from the SDSS to find that SF is indeed enhanced by tidal interactions, and that this enhancement depends strongly on the separation in distance between the galaxies, but, surprisingly, only very weakly on the relative luminosity of the companions. This latter fact is of interest because we will, below, use a lower limit in difference of 3\,mag to identify companions to the H$\alpha$GS sample galaxies. As a second example, Ellison et al. (2008) select 1716 galaxies with close companions (pair velocity difference $\Delta\,v<500$\,km\,s$^{-1}$, pair separation $r<80\,h^{-1}_{70}$\,kpc, and mass ratio between 1 and 10) from the SDSS and find an enhancement of the SFR for close pairs ($r<30-40\,h^{-1}_{70}$\,kpc), stronger for pairs of galaxies of approximately equal mass. 

SDSS-selected samples have the obvious advantage of larger size over our H$\alpha$GS sample. This implies that either very large numbers of galaxies can be studied, such as in the work of Li et al. (2008), or that rather strict selection criteria can still yield samples of large size, as in Woods \& Geller (2007) or Ellison et al. (2008). But use of \ha\ imaging of local galaxies such as in H$\alpha$GS has at least two important advantages that can be offset against the disadvantage of smaller sample size. These are, first, that H$\alpha$GS yields measures of the SFR and \ha\ EW integrated over the whole galaxy, and, second, that we can trace the detailed behavior of these quantities within galaxies, using the images (with a typical resolution of 1.5\,arcsec or 150\,pc), or using radial profiles (see Sect.~4.3). In contrast, SDSS studies are based on the fibre-fed spectra obtained in the SDSS survey, which are limited to a single 3\,arcsec aperture centered on a target galaxy. This means that the SFR information deduced from the SDSS is that of the central area only for nearby galaxies, and it thus ignores SF activity in the disk (a particularly important point in late-type galaxies---see the H$\alpha$GS paper by James, Bretherton, \& Knapen 2008a [Paper~VII]). Measurements for individual galaxies may also be influenced by non-stellar emission from the nuclear region. In addition, whereas the \ha GS galaxies are all within a redshift of 0.01 ($v_{\rm sys}<3000$\,km\,s$^{-1}$; see Sect. 2.1), the galaxies considered by Ellison et al. (2008) and Li et al. (2008) span a much larger range in distance (up to $z\sim0.16$ and 0.30, respectively). As galaxies with very high SFRs ($>10\,M_{\odot}$\,yr$^{-1}$) are extremely rare in at least the local Universe, they are much more prominent in the SDSS-based studies than in ours. We quantify this further in Section~2.1.

The advantages of using \ha\ imaging to study the SF properties of local galaxies must be offset by the main disadvantage which is that the \ha\ line is located in a region of the spectrum where dust attenuation is not negligible. 
Comparing SFR values derived from the \ha\ line with extinction-free measures in the radio continuum (e.g., van der Hulst et al. 1988) or far-IR (e.g., Rosa-Gonz\'alez et al. 2002) wavelength domains yields factors of a few by which the \ha\ line alone will underestimate the true SFR. As reviewed by Calzetti (2008), brighter galaxies are more extinguished. Even though most of our sample galaxies will be affected less because they are faint dwarfs and irregulars, the effect of dust on our SFR determinations is a caveat worth noting. Our main conclusions, however, are all based on comparisons of sub-samples and/or individual galaxies selected from the same overall \ha GS data set, rather than on the absolute values of the SFRs, thus extenuating any dust effects. 

In the present paper, we first discuss the origin of our observational data (Sect.~2) and how exactly we define whether a sample galaxy has a close companion (Sect.~3). In Sect.~4 we present the measured values of the SFR and \ha\ EW in the sub-samples of galaxies with and without a close companion, and discuss possible sample biases, the radial dependence of the SFR and EW, and the gas content. In Sect.~5, we highlight the interacting and merging galaxies, and the infrared-luminous galaxies in our sample. In Sect.~6, we discuss those galaxies which have the highest values of SFR, EW, and central and overall SFR per unit area, and those that have the lowest gas depletion timescales. We use these results to discuss critically the possible definitions of the term ``starburst'' in Sect.~7, where we also address the implications of our results for the evolution of galaxies, and in particular the effects of close companions and interactions on the SF in galaxies. We summarise our conclusions in Sect.~8. 


\section{Sample selection, observations, and data treatment}\label{data}

\subsection{H$\alpha$ GS data}

All optical imaging data used in this paper was taken as part of the H$\alpha$GS, which is a survey of 334 nearby galaxies, imaged in the \ha\ line and in $R$-band continuum with the 1\,m Jacobus Kapteyn Telescope (JKT) on La Palma. The narrow-band filters used were wide enough that what we refer to here as \ha\ emission is in fact a combination of that in the \ha\ and neighboring [N{\sc ii}] lines, although in Paper~I we corrected the derived SFR and EW width values for this. 

The \Ha GS sample was selected from the
Uppsala Galaxy Catalogue (UGC; Nilson 1973).  
The \Ha GS selection criteria are described in detail in Paper I, but
can be summarised as follows: type S0/a ($T=0$) or later; diameters
between 1.7 and 6.0 arcmin; recession velocities less than
3000~km\,s$^{-1}$, and declinations greater than 
$-2.5^{\rm o}$. As discussed in more detail by James et al. (2008b; hereafter Paper~IV), the UGC provides an optically-selected
sample with excellent completeness over the luminosity and distance range 
of the galaxies considered. 

To quantify this further, in particular to confirm that our sample is not biased against galaxies with high SFRs, we calculate how many such galaxies might be expected within the \ha GS. We used the \ha\ luminosity functions (LFs) in Ly et al. (2007, their Fig. ~10) and Shioya et al. 
(2008, their Fig.~4) to estimate the expected numbers of galaxies of different SFR within 
the \ha GS volume.  We first chose to check galaxies with SFR$> 30$\,$M_\odot$\,yr$^{-1}$, and find that one might expect 
$2 - 8 \times 10^{-5}$\,galaxies\,dex$^{-1}$\,Mpc$^{-3}$ (using the \ha--SFR conversion factor of $7.9\times10^{-42}$
from Kennicutt et al. 1994; the range in the result stems from the different lines fitted 
to the above figures).  Selecting the SFR range $20-40$\,$M_\odot$\,yr$^{-1}$, i.e. 0.3~dex in SFR, the predicted number density is $0.6 - 2.4 \times 10^{-5}$\,galaxies\,dex$^{-1}$\,Mpc$^{-3}$. 

The \ha GS survey volume is $\pi \times 40^3 \times (1.39/4)$\,Mpc$^3 = 7 \times 10^4$\,Mpc$^3$ (Paper~IV)
but we only sample approximately a third of this so the effective volume is $2.3 \times 10^4$\,Mpc$^3$.

In the \ha GS sample, we should thus expect only $0.14 - 0.55$ galaxies with a SFR of $20-40$\,$M_\odot$\,yr$^{-1}$  (and far fewer above this range).
In fact we have none, but we have three  in the range $10-20$\,$M_\odot$\,yr$^{-1}$, which is exactly in line 
with the above predictions.

We additionally checked around the ``knee" in the LF, which is where the 
dominant contributors to the total SFR density lie.  This is about 4.4\,$M_\odot$\,yr$^{-1}$, so we derived the expected numbers in the range $2 - 10$\,$M_\odot$\,yr$^{-1}$, 
getting predictions of 26 (using the Ly et al. 2007 LF) and 44 (Shioya et al. 2008), whereas we in fact have 33 
\ha GS galaxies in this range. We thus conclude that the {\it \ha GS does
accurately sample the star-forming galaxy population}. HaGS is not just a 
survey of low-luminosity galaxies, and the luminous and strongly star-forming galaxies which one finds in many studies based on larger volume surveys like the SDSS really are very rare.  To illustrate this further, we point out 
that the two most famous and well studied luminous mergers, Arp~220 and 
NGC~6240, lie at ~80 and 100~Mpc distance, respectively, well outside our 
survey volume.

Of the 334 galaxies reported in Paper~I, we use 327 in this paper (as explained in Paper~IV, the remaining seven were serendipitously observed). The values for the SFR, EW, and distance for each sample galaxy were taken from Table~3 of Paper I. In that paper, reliable distances were calculated using a Virgocentric model for the local Hubble flow. Absolute magnitudes were calculated using the distance and the $B_{\rm T}$ magnitude values from NED. The criteria used for determining which galaxies have close companions or are interacting are described below, in Section~\ref{definitions}.

We use the primary H$\alpha$GS data products SFR (in units of solar masses per year) and EW (in units of nm). In addition, we use a number which we call the gas depletion timescale, $\tau$, and which we obtain by simply dividing the total amount of gas in a galaxy (derived from the total H{\sc i} content, see Sect.~2.2), in units of solar masses, by the SFR. Thus, $\tau$ would be the time needed to deplete the gas reservoir in a galaxy if the SF were to continue at a constant rate equal to the current SFR. Below, and especially in Sect.~7.1, we will discuss the validity of using this measure to relate the SFR in a galaxy to its available fuel, as has been proposed in the literature in relation to defining starbursts.

In Sect.~4.3, we use the radial profiles of SF and \ha\ EW we derived in Paper~VII. These are mean normalised light profiles, which relate the flux within elliptical annuli, but, unlike in traditional surface brightness or equivalent profiles, {\it not} corrected for the increasing  area in the annuli as the apertures grow. This procedure yields curves in which the area under a profile within a given range of radius is directly proportional to the amount of light contributed to the total luminosity of the galaxy. These curves are normalised, both in radial scale (in units of the 24\,mag per square arcsec isophotal semi-major axis of the galaxy) and in intensity (where the area under the curve is set to unity). The profiles are thus shape functions, indicating the radial distribution of flux.

\subsection{Atomic gas data}

Data on the atomic gas for the sample galaxies were collected from the HyperLeda database (Paturel et al. 2003), which contains data for all but six of our galaxies. Using the distance (Paper~I) we transformed the 
H{\sc i}  flux, $f$, given in HyperLeda, to units of solar masses using the standard formula $M_{\rm HI}=235.6\,D^2f$. For three very close pairs of galaxies, and due to the beam sizes used in the original surveys, only one H{\sc i}  flux measurement is given. We made no attempt to separate such measurements, because our goals for the use of H{\sc i}  data are purely statistical, as described in Section~\ref{hiresults}.

In order to calculate the gas depletion timescales we use later in this paper, we first multiply the atomic gas mass by a factor 2.3 to include an estimate of the molecular part of the hydrogen gas reservoir as well as the helium (analogous to H$\alpha$GS Paper~V by James, Prescott \& Baldry 2008c, who in turn took the factor of 2.3 from Meurer et al. 2006), and divide this by a factor of $(1-0.4)=0.6$, where 0.4 is assumed to be the mass fraction of the stars formed which is recycled to the interstellar medium (see paper~V). We finally divide the resulting number (in Solar masses) by the SFR (in Solar masses per year) to get the gas depletion timescale, $\tau$, in units of years. 

We make an independent estimate of 
gas masses by following the suggestion of Bigiel et al. (2008) that 
molecular masses scale roughly linearly with SFR, adding the \hi\ mass
(corrected for \hi\ self-absorption) and multiplying the total by 1.35 to 
account for helium. This results in gas masses which are, after the correction for the recycling, smaller than our original
estimates by 45\% (probably a reasonable reflection of the systematic 
uncertainties in absolute gas masses), with a standard deviation of 14\%. 
This latter figure represents the relative uncertainty in gas depletion 
time from galaxy to galaxy, and does not substantially affect our 
comparisons of this parameter between galaxies, or between subsets of our sample.


\section{Presence of companion galaxies}\label{definitions}

To decide whether a galaxy has a close companion or not, we used a number of criteria that have evolved from those used by Schmitt (2001), Laine et al. (2002), and Knapen (2005), and that were used previously by us in Knapen et al. (2006). Using the HyperLEDA and NASA-IPAC extragalactic databases, we search the neighborhood of all H$\alpha$GS galaxies for companion galaxies that are not only in the close vicinity, but also have sufficient mass so we can expect them to have some gravitational impact on the H$\alpha$GS galaxy. Such close companions must satisfy each of the following three criteria. They must (1) be within a radius of five times the diameter of the galaxy under consideration, or $r_{\rm comp}<5\times D_{25}$, where $D_{25}$ is taken from the RC3; (2) have a recession velocity within a range of $\pm$200\,km\,s$^{-1}$ of the sample galaxy; and (3) be not more than three magnitudes fainter.

\begin{figure*}
\begin{center}
\includegraphics[width=\textwidth]{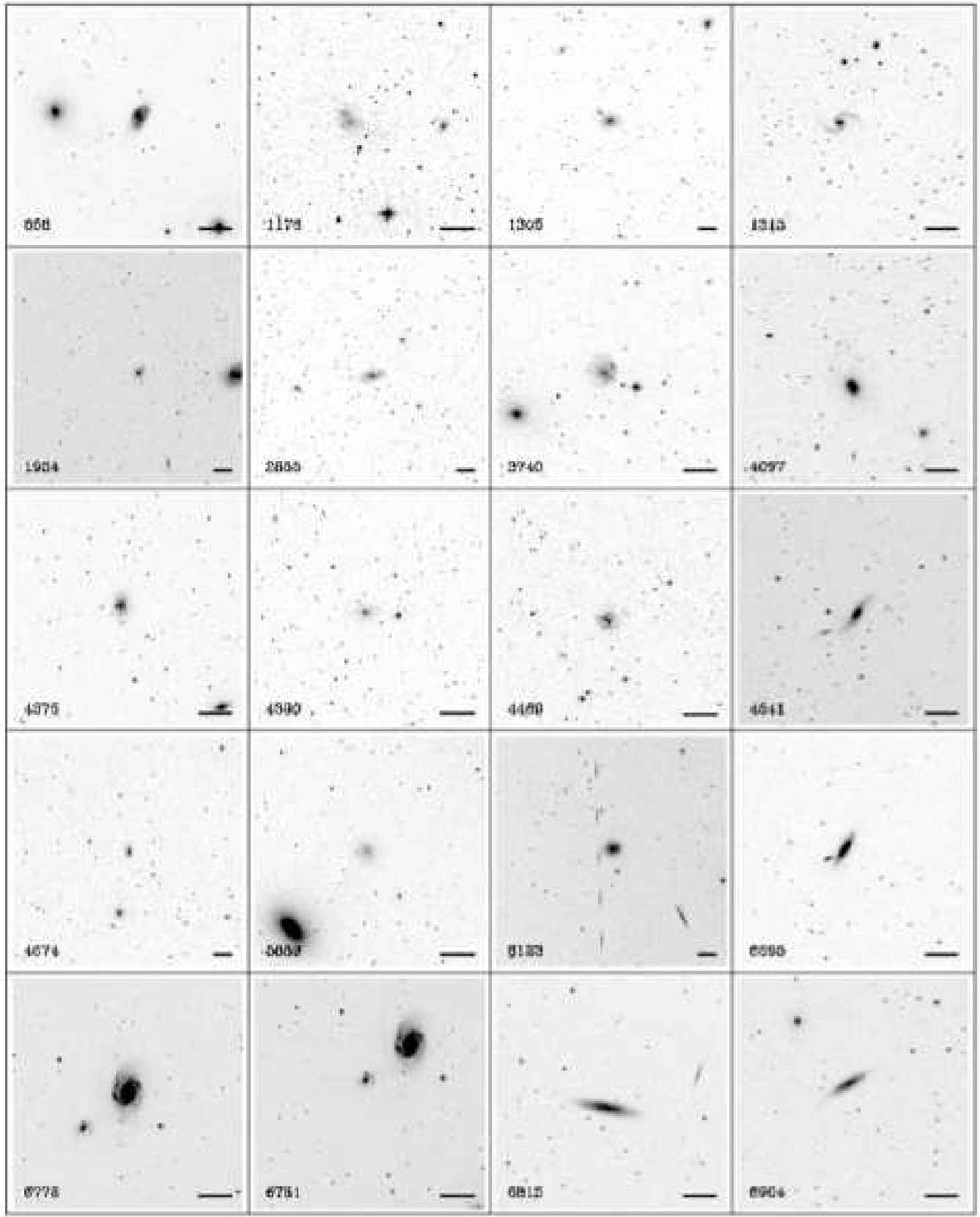}
\end{center}
\caption{DSS images as obtained from the NED for those H$\alpha$GS galaxies that have a close companion. The area shown varies so that the companion is included in each image. The H$\alpha$GS galaxy is in all cases in the center of the image. Where both galaxies in a pair are in H$\alpha$GS, two images have been included. Galaxies are identified by their UGC number in the bottom left corner of each image. The scale bar in the bottom right corner indicates 2~arcmin on the sky.
}
\end{figure*}

\setcounter{figure}{0}

\begin{figure*}
\begin{center}
\includegraphics[width=\textwidth]{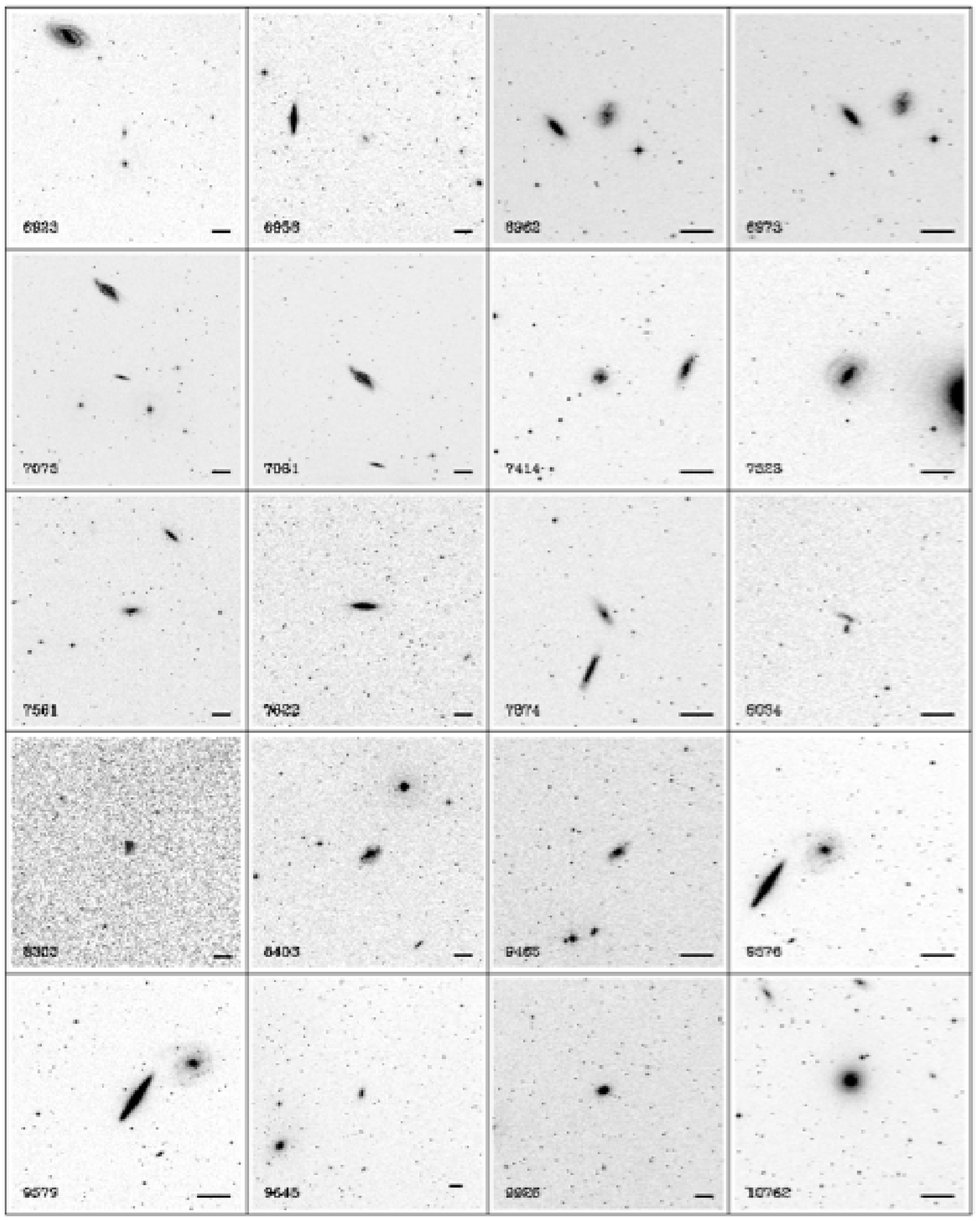}
\end{center}
\caption{(ctd.)
}
\end{figure*}

\setcounter{figure}{0}

\begin{figure*}
\begin{center}
\includegraphics[width=\textwidth]{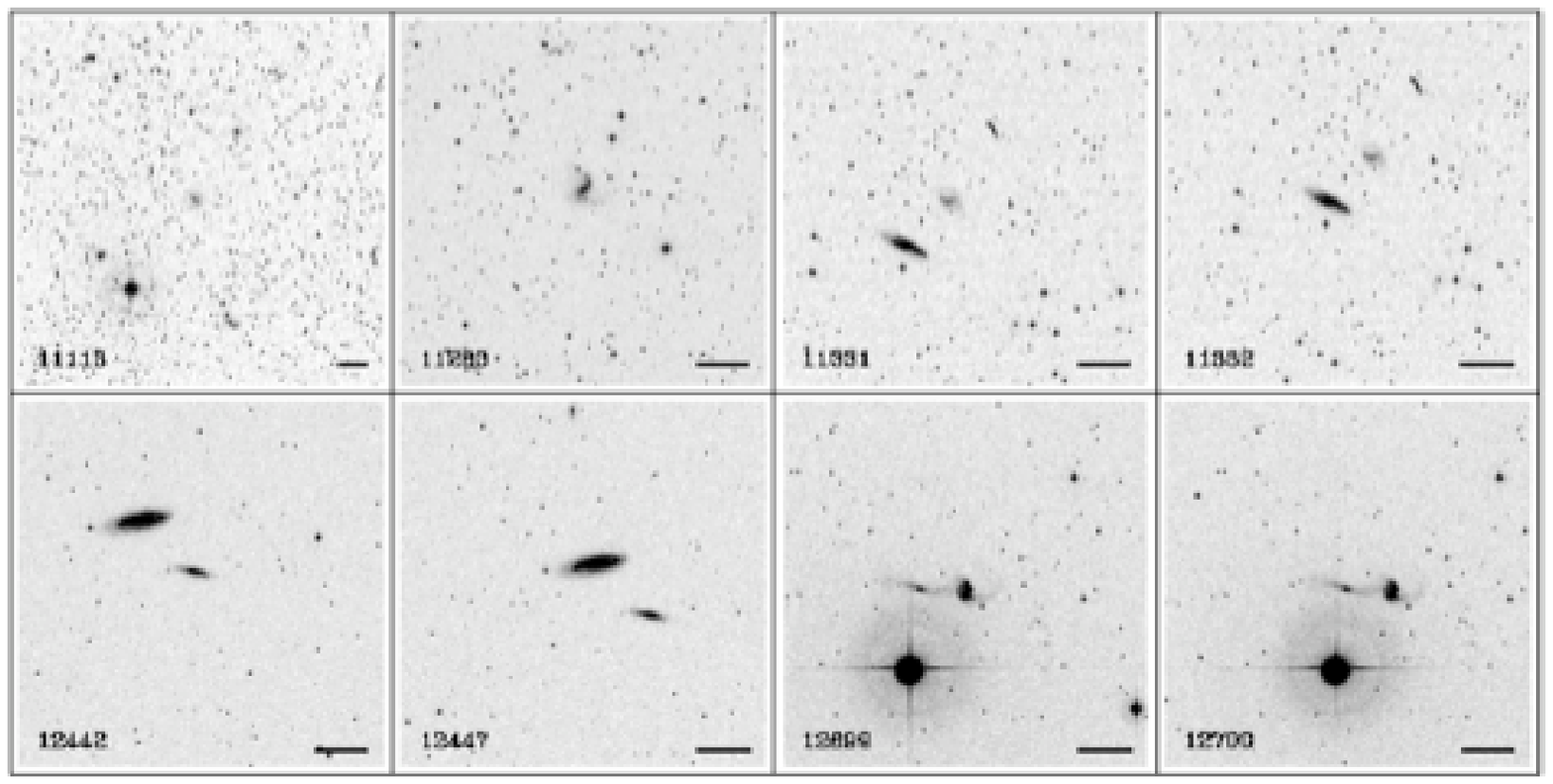}
\end{center}
\caption{(ctd.)
}
\end{figure*}

We find that 48 of our 327 H$\alpha$GS galaxies (some 15\%) have a close companion following these criteria, and digital sky survey images of these 48 galaxies and their companions are shown in Fig.~1. The criteria should be robust and allow us to make a complete search of close companions, except possibly in the case of very faint H$\alpha$GS galaxies. For those, companions which are another 3\,mag fainter will not all be catalogued, and may thus be missed. In order to check that this does not jeopardise our results, we have made a number of robustness tests, for instance by considering only the bright half of our H$\alpha$GS sample. This is described in Section~\ref{biases}.

The low number of galaxies with close companions does not allow us to make the criteria even more stringent while maintaining a statistically meaningful interpretation of our results. Because of the interest of studying companions closer than the limits used in our criteria, though, we discuss below the small number of galaxies which can in fact be considered to be interacting or merging (Section~5). But we will first, in Section~\ref{results}, explore the effect of close companions on the SFR and \ha\ EW in local galaxies.


\section{Results: Close companions and massive star formation}\label{results}

\subsection{Star formation rates and equivalent widths}

\begin{table}
\begin{center}  
\begin{tabular}{cccccc}
\hline
\hline
Type & Ratio & Std. Error & $N_{\rm cl. comp.}$ & $N_{\rm tot}$ & KS-test $P$\\ 
\hline
{\bf SFR}: &&&&&\\
All & 1.85 & 0.30 & 48 & 279 & 0.008\\
0--2 & 0.52 & 0.12 & 7 & 29 & ND\\
3--8 & 2.23 & 0.39 & 33 & 141 & 0.006\\
9--10 & 1.43 & 0.60 & 8 & 109 & ND\\
\hline
{\bf EW}: &&&&&\\
All & 1.16 & 0.11 & 48 & 279 & 0.005\\
0--2 & 0.85 & 0.24 & 7 & 29 & ND\\
3--8 & 1.31 & 0.14 & 33 & 141 & 0.001\\
9--10 & 0.82 & 0.21 & 8 & 109 & ND\\
\hline
\end{tabular}
\caption[]{Mean SFR and \ha\ EW of galaxies with close companions normalised to the mean of all galaxies in the (sub-)sample without a close companion. This is shown for all galaxies in the sample, and separately for galaxies of certain morphological type groups (col.~1). The ratio is shown in col.~2 and the standard error on this number in col.~3. A ratio larger than unity means that the SFR/EW is enhanced in galaxies with a close companion as compared to those without. The number of galaxies in each sub-sample with and without a close companion is shown in cols.~ 4 and 5, respectively. Results ($P$-value) from a KS test which indicates whether differences are statistically significant are shown in col.~6, where `ND' means no difference ($P>0.05$).}
\label{SFR+EW}
\end{center}
\end{table}

The question we wish to answer here is whether the presence of close companions enhances the SFR and/or the EW of a galaxy. 
We compare the mean SFR and EW values of galaxies with a close companion to those of galaxies without one. For each of the former group of galaxies, we divide its SFR by the 
mean SFR for the H$\alpha$GS galaxies without a close companion; we then determine the statistical parameters
(mean, standard deviation, standard error) on those ratioed SFRs, and analogously for the EWs. If galaxies with close companions were to have enhanced massive SF (higher SFR and/or EW), these values should be higher than unity. If, on the other hand, massive SF were suppressed in galaxies with close companions as compared to those without, the value would be smaller than one. The results are shown in Table~\ref{SFR+EW} for the complete sample of 327 H$\alpha$GS galaxies, and for three smaller sub-samples of different morphological type. 

Table~\ref{SFR+EW} shows, firstly, that the presence of a close companion in galaxies is accompanied by a significantly higher SFR, but by only a marginally higher EW. This effect is particularly driven by those galaxies of intermediate morphological types. We interpret this as due to an enhanced $R$-band continuum luminosity, which cancels the increased \ha\ emission due to the higher SFR. This implies that the SFR, which is increased in relation to the presence of the close companion, must have been enhanced over a long enough timescale to allow the formed stars to evolve to a stage where they emit significant quantities of light in the $R$-band, or a few hundred million years.

Secondly, considering galaxies of different morphological types, we find that the SFR and EW are suppressed, but not significantly so, in early-type galaxies with a close companion. The EW is also suppressed (though not significantly so) for those of the latest types. The SFR of the latest-type galaxies with close companions is enhanced, but again this result is not statistically significant. Of more interest is our result  that the SFR and, to a lesser extent, the EW of mid-type (3--8, or Sb to Sdm) galaxies is enhanced in the presence of  a close companion. This result, as judged from the standard error and confirmed by a Kolmogorov-Smirnov (KS) test (see Table~\ref{SFR+EW}), is marginally significant, and may well be related to the results from numerical modelling by Mihos \& Hernquist (1994; see also Springel  2000) that the presence of a central massive bulge component stabilizes the disk, and delays gas inflow to and star-forming activity in the central region until the final stages of a merger.

\begin{figure}
\begin{center}
\includegraphics[width=0.45\textwidth]{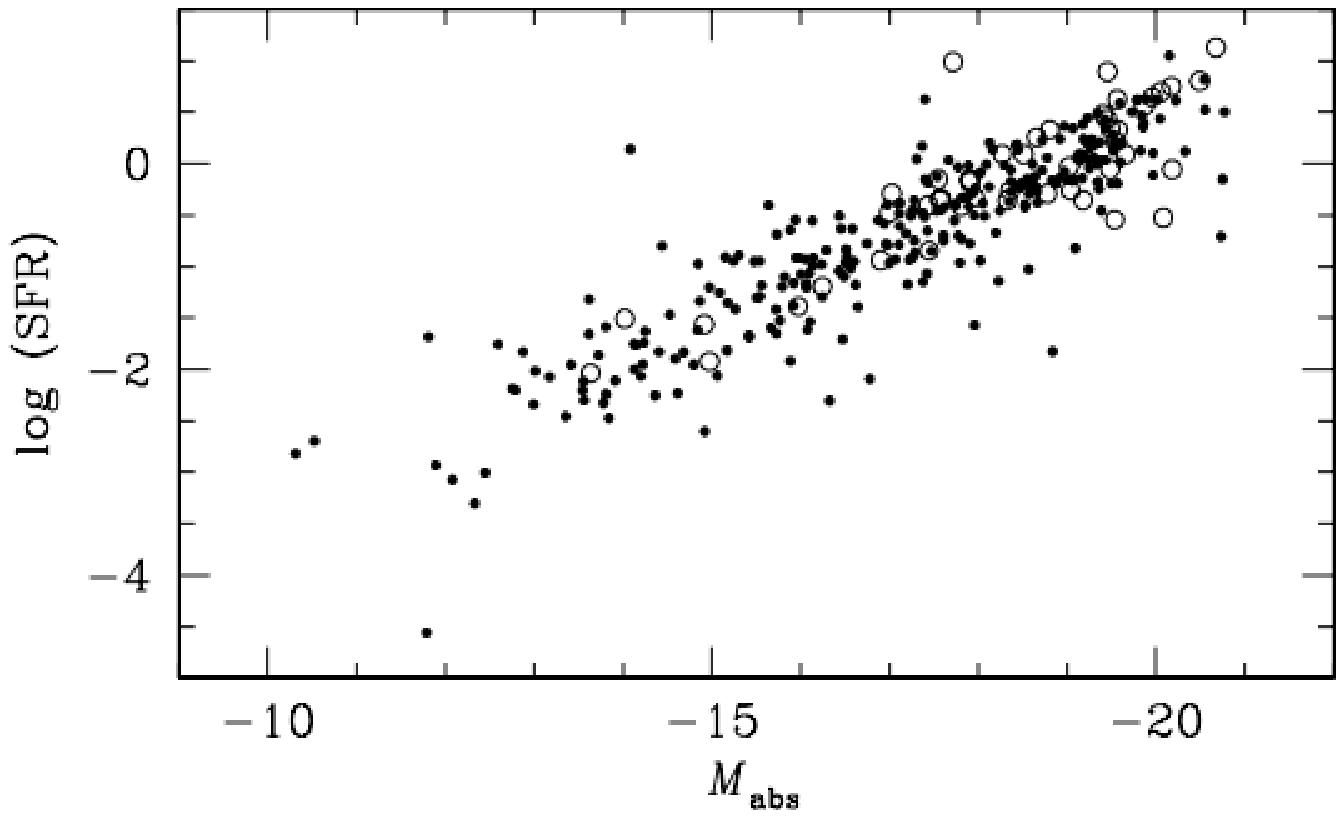}
\end{center}
\caption{Logarithm of the SFR, in units of solar masses per year, as a function of the absolute magnitude of the sample galaxies. \ha GS galaxies with a close companion are identified as open circles, whereas those without a close companion are marked as filled dots.
}
\label{figsfrmabs}
\end{figure}

\begin{figure}
\begin{center}
\includegraphics[width=0.45\textwidth]{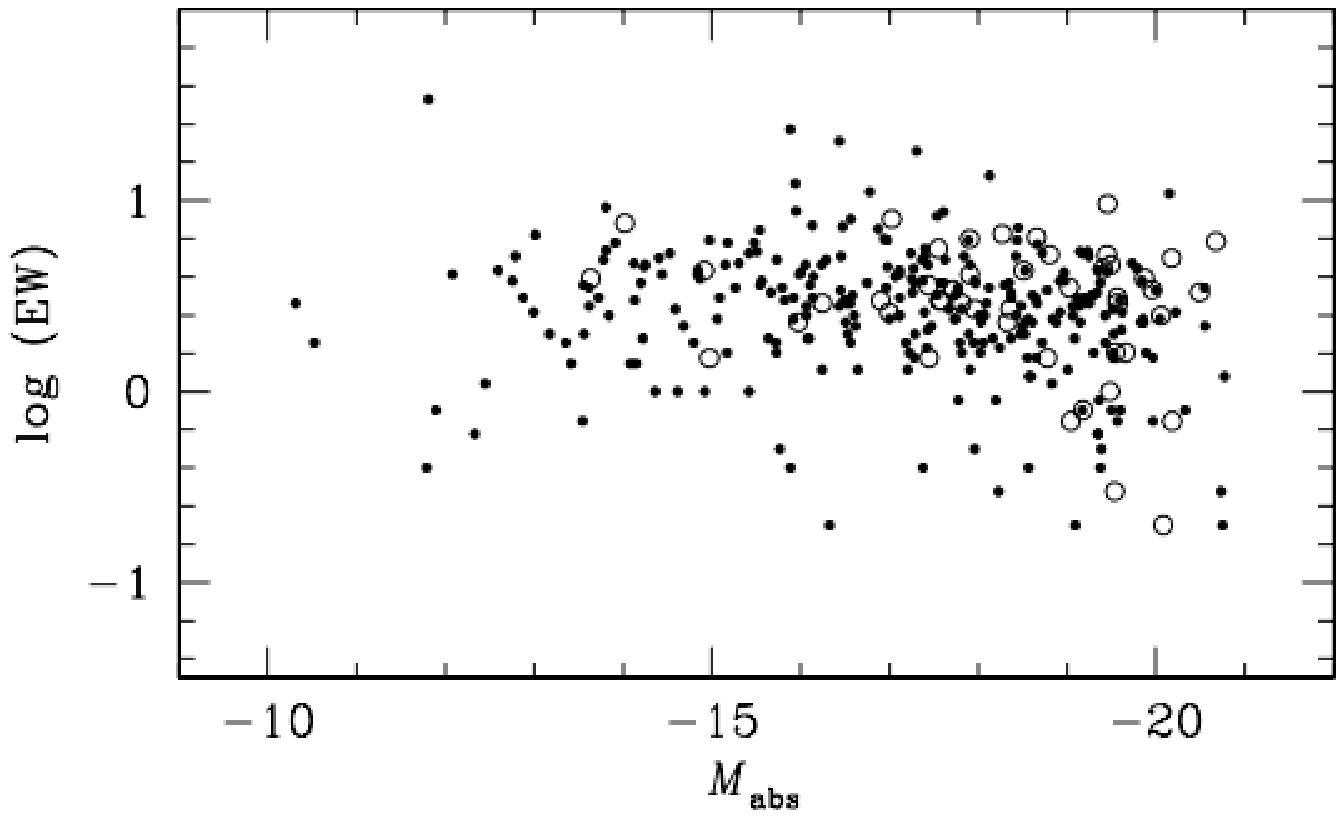}
\end{center}
\caption{As Fig.~\ref{figsfrmabs}, but now for the EW (in units of nm).
}
\label{figewmabs}
\end{figure}

Figure~\ref{figsfrmabs} shows the SFR in solar masses per year, as derived from our \ha\ imaging and uncorrected for extinction, as a function of the absolute magnitude of our \ha GS galaxies, marking those with a close companion as open circles, and those without, as filled dots. This plot is analogous to the one shown in Fig.~6 of Lee et al. (2008), and we refer the reader to their paper for a more general discussion of the observed rise of SFR with absolute magnitude, hence mass of the galaxy, and of further implications of this diagram especially for the dwarf galaxy population. 

The figure shows how the galaxies with a close companion are located preferentially on the right hand side of the diagram, indicating that they tend to be more massive than the median across the whole sample, but that their SFRs are not enhanced with respect to the median of the distribution at their absolute magnitude. This provides a qualitative manner of visualising our main result: the SFR of the galaxies with a close companion is enhanced with respect to the parent sample mainly because they are big (luminous, massive) galaxies. 

Since the EW measures in some way the SFR normalized by the underlying galaxy continuum emission, or mass, the gradual rise of SFR with $M$ as seen in the diagram is divided out, and the EW of those galaxies with a close companion is hardly enhanced at all with respect to those that do not have a close companion. This is shown graphically in Fig.~\ref{figewmabs}, where the EW is seen to remain constant across the whole range sampled, of some 11 magnitudes in $M$. In Paper~I we already showed such a figure for the \ha GS galaxies, and noted that the effect of constant EW with magnitude had  been published before in the literature, but not over such a large range (see, e.g., Kennicutt \& Kent 1983). Lee et al. (2007) present a similar figure as based on data from the 11HUGS (Kennicutt et al. 2008), and code the datapoints by morphological type. The downturn in EW at the high-$M$ end of the distribution is due to massive, early-type, spirals with little current massive SF but large masses, but a contribution of dust extinction in these high-mass galaxies cannot be excluded---this would diminish the SFR and thus the EW.  From the numbers given above, we already know that the EWs of those galaxies with and without a close companion are comparable in the median, and this result is confirmed nicely in Fig.~\ref{figewmabs}, where the former are sprinkled throughout the diagram (though with a slight preference to the right, high-mass, side of the diagram, as we noted before).

\begin{figure}
\begin{center}
\includegraphics[width=0.45\textwidth]{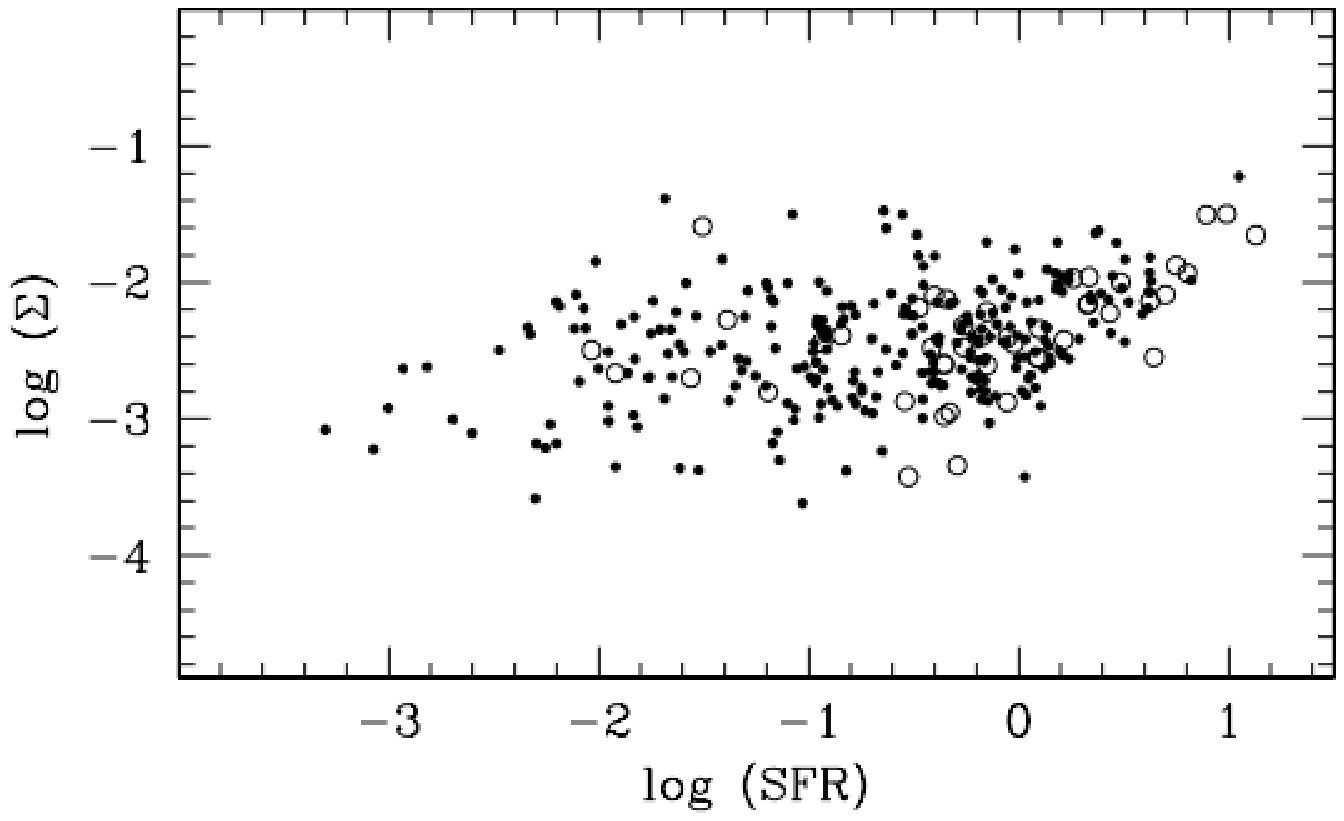}
\end{center}
\caption{Logarithm of the SFR per unit area, $\Sigma$, in units of solar masses per year per square kiloparsec, as a function of the logarithm of the SFR, in units of solar masses per year, for all the sample galaxies. As in Fig.~\ref{figsfrmabs}, those \ha GS galaxies with a close companion are identified as open circles, whereas those without a close companion are marked as filled dots.
}
\label{figsigmasfr}
\end{figure}

Figure~\ref{figsigmasfr} shows how the SFR per unit area, $\Sigma$, as calculated by normalizing the SFR by the area of the galaxy within the radius $r_{24}$ (see also Sect.~6.3), behaves as a function of the SFR. Kennicutt et al. (2005) use such a diagram to locate starburst galaxies, but our emphasis here is again to show the relative location of those galaxies with and without a close companion. We see once again that the galaxies with a close companion are inconspicuous, except for the dozen or so galaxies located in the top right corner of the diagram, and which are mostly there because of their high SFR. As we have noted before, galaxies with close companions tend to be in the right hand side of the diagram, with many showing SFRs above the median value for the sample. Their $\Sigma$ values, though, are not different on average from those of the galaxies without a close companion. 

\subsection{Possible sample biases}\label{biases}

Our method is in principle susceptible to a number of systematic issues, and in this Section we describe a number of tests performed to ascertain to what degree these issues might influence our final results. The first effect we consider is that of apparent brightness of the sample galaxies. For the faintest H$\alpha$GS galaxies, it is possible that our criterion that companion galaxies need to be at most 3\,mag fainter than the sample galaxies fails, in the sense that the faintest possible companions will be below the brightness cut-off level of the main galaxy catalogues. This would mean that a faint galaxy can have a very faint companion, but the latter would not be registered by us. 

The second effect is that of distance. For nearby galaxies, fainter companions will generally be included in the catalogues (unless they are very faint, see preceding paragraph) but for distant ones this may be less true.

\begin{table}
\begin{center}  
\begin{tabular}{ccc|ccc}
\hline
\hline
Sub-sample & \multicolumn{2}{c}{SFR} & \multicolumn{2}{c}{EW} & N\\ 
 & mean & error & mean & error & $N$ \\
\hline
Near & 0.58 & 0.21 & 1.15 & 0.17 & 12\\
Far & 1.69 & 0.27 & 1.20 & 0.14 & 36\\
Faint & 1.78 & 0.42 & 1.14 & 0.16 & 17\\
Bright & 1.88 & 0.37 & 1.25 & 0.16 & 31\\
\hline
\end{tabular}
\caption[]{Mean SFR and \ha\ EW of galaxies of those galaxies with a close companion normalised by that of those without a companion, for four equal-sized sub-samples, namely those of H$\alpha$GS galaxies that are close vs far, and faint vs bright. }
\label{halves}
\end{center}
\end{table}

Because any specific test will diminish the number of galaxies in the sub-sample considered, we opted for a series of simple tests which maintains to the maximum the number of galaxies in the sub-samples to be tested. We thus split the overall sample in two halves with equal numbers of galaxies (163 and 164), taking as criteria apparent magnitude (faint half vs bright half) and distance (close vs far). The results of these tests on the mean SFR and EW  of galaxies with and without a close companion are listed in Table~\ref{halves}.

The values in Table~\ref{halves} are the mean and standard error of SFR and \ha\ EW for those galaxies with a close companion as compared to those without. They were calculated by taking the SFR and EW values for each of the former galaxies, and dividing them by the mean value for galaxies of their type, and in their half of the sample (near vs far, bright vs faint).  The large value for the ``faint'' SFR is caused by two Im galaxies with a close companion in the faint half which have a SFR six and eight times higher than the mean over all Im galaxies in the faint half. The EW appears to be more robust than the SFR against such large excursions. Using a series of KS tests to ascertain whether any of the differences, between near and far, or between bright and faint, are statistically meaningful, we confirm the impression one gains from the errors cited in the Table that this is not the case---none of the differences is significant to the 90\% level.

The conclusion we can draw from this test, as summarised in Table~\ref{halves}, is that neither distance nor brightness significantly affect  the final results of our study. The numbers for the SFR are consistent in the far/bright sub-samples, but not in the near/faint ones, presumably because of the small numbers of galaxies. The EW values can be considered equal for the four sub-samples under consideration.

\subsection{Radial dependence within galaxies}

In this Section we describe the results of a more detailed analysis of the \Ha\ emission from \Ha GS galaxies with and without a close companion, namely how their \Ha\ emission is distributed radially within the galaxies. We do this specifically to shed light on the question of whether the presence of a close companion causes a central concentration of the massive SF activity, as identified, in more extreme cases, with (circum-)nuclear starbursts. For instance, Bushouse (1987) found that the star-forming activity in interacting galaxies is indeed concentrated near their nuclei, as expected as a result of inflow either by the interaction itself, or by bars induced by the interaction. 

First, we consider the SFR and EW of the inner as compared to the outer halves of each \Ha GS galaxy, comparing those with and without a close companion.

Of the 48 galaxies with close companions, H$\alpha$GS provides usable profiles for 45.  Of these 45, 22 galaxies, or  $49\pm13\%$, have a higher EW inside the half-light radius than outside.  For the rest of the H$\alpha$GS sample, those galaxies without a close companion, this is the case for 100 out of 269, or $37\pm4\%$.

We thus find a trend for enhanced central SF in galaxies with close companions, although this trend is not statistically significant. To explore whether the study of smaller sub-samples, or of galaxies of specific morphological type, can shine further light on this, we perform a more detailed study of the radial distribution of SF.

As  a second test, we thus study the radial flux profiles, as described in detail in Paper~VII and as summarised in Sect.~2 of the current paper, of the \Ha\ emission of all \Ha GS galaxies of a particular morphological class with and without a close companion. We express the results of this test as the fraction of galaxies in which the Ha EW is higher inside $0.2\times r_{24}$ than within $1.0\times r_{24}$. For the complete sample, we find that $36\pm11$\% (16/45) of the galaxies with a close companion have such centrally concentrated \ha\ EW, vs $30\pm4$\% (80/268) or those without a close companion. For galaxies of morphological types $T=3-8$ the numbers are $31\pm13$\% (10/32) and $21\pm5$\% (29/138), respectively, and for galaxies of type $T=3$, $50\pm49$\% (3/6) and $41\pm22$\% (7/17).

\begin{figure}
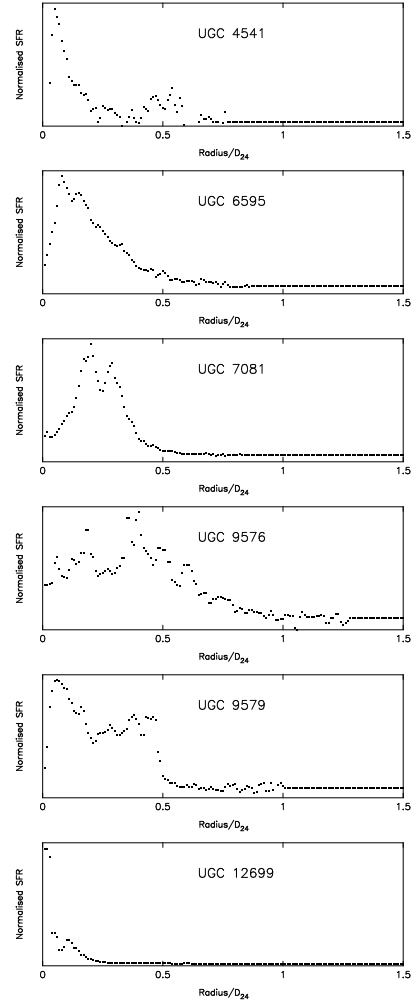

\begin{center}
\rotatebox{-90}{
\includegraphics[width=2.1cm]{u4541profn.ps}}
\rotatebox{-90}{
\includegraphics[width=2.1cm]{u6595profn.ps}}
\rotatebox{-90}{
\includegraphics[width=2.1cm]{u7081profn.ps}}
\rotatebox{-90}{
\includegraphics[width=2.1cm]{u9576profn.ps}}
\rotatebox{-90}{
\includegraphics[width=2.1cm]{u9579profn.ps}}
\rotatebox{-90}{
\includegraphics[width=2.1cm]{u12699profn.ps}}
\end{center}
\caption{
Normalised \Ha\ profiles for interacting/merging systems: top to bottom
UGC~4541, UGC~6595, UGC~7081, UGC~9576, UGC~9579 and UGC~12699.
}
\label{figprofiles}
\end{figure}

We find that in all these sub-samples there is a small increase in the central SF which is linked to the presence of a close companion, but in no case is this significant, even at the $1\sigma$ level. The $T=3$ galaxies (whether they have a close companion or not) show more central concentration than other galaxies---this is the effect of the $T=3$ barred galaxies discussed in more detail in Paper~VII. As an example of the kind of profiles we find for the interacting/merging galaxies identified here (Sect.~5), we show in Fig.~\ref{figprofiles} the radial flux profiles for six of these. The profiles can be compared with the average ones for the relevant morphological type as published in Paper~VII, but apart from a central concentration in a few galaxies (such as UGC~12699) there are no obvious significant differences.

Each of our two tests thus reveals that there is a statistical tendency for galaxies with a close companion to have a higher degree of central concentration of their SF than galaxies without a close companion, but none of the tests returns a difference between these samples which is significant---not even at the level of $1\sigma$. This may well be because most of our close companion galaxies are far from mergers, and because it is in the advanced merger stage where one might expect enhanced SF in the central region. Another reason may be that in galaxies which are not mergers, bars are the most efficient drivers of gaseous inflow, and thus of central SF. But as two thirds of galaxies have bars anyway, any incremental effect due to interactive effects can only be limited. In addition, bar triggering by interactive events only happens primarily in prograde, and less in retrograde, encounters (e.g., Gerin et al. 1990; Berentzen et al. 2004; Romano-D\'\i az et al. 2008), limiting any effect even more. We conclude that there is no evidence for the presence of a close companion leading to centrally concentrated SF.

\subsection{The role of atomic gas}\label{hiresults}

\begin{table}
\begin{center}  
\begin{tabular}{cccccccc}
\hline
\hline
& \multicolumn{4}{c}{Whole sample} & \multicolumn{3}{c}{With Close Comp.}\\
$T$ & $N$ & $\tau$ & $M_{\rm HI}$ & $M_{\rm HI}/L_R$ & $N$ & $M_{\rm HI}$ & $M_{\rm HI}/L_R$  \\
  &  & $10^9$\,yr & $10^9\,M_{\odot}$ & $M_{\odot}/L_{\odot}$ &  & $10^9\,M_{\odot}$ & $M_{\odot}/L_{\odot}$ \\
\hline
0  & 13 & 6.9 & 0.88 & 0.436 & 1 & 0.85 & 0.075\\
1  & 12 & 11 & 2.23 & 0.325 & 3 & 1.60 & 0.193\\
2  &  7 & 4.8 & 1.68 & 0.303 & 1 & 0.42 & 0.245\\
3  & 26 & 4.3 & 1.91 & 0.302 & 7 & 1.93 & 0.240\\
4  & 24 & 4.5 & 2.70 & 0.358 & 5 & 3.26 & 0.519\\
5  & 36 & 4.0 & 2.29 & 0.474 & 9 & 3.46 & 0.425\\
6  & 31 & 9.3 & 1.80 & 0.825 & 4 & 2.02 & 0.456\\
7  & 28 & 11 & 1.42 & 0.935 & 4 & 2.70 & 0.942\\
8  & 29 & 9.6 & 1.07 & 0.839 & 4 & 1.05 & 0.588\\
9  & 41 & 22 & 0.79 & 2.086 & 3 & 0.85 & 2.283\\
10 & 73 & 21 & 0.50 & 2.571 & 3 & 0.95 & 1.597\\
\hline
\end{tabular}
\caption[]{H{\sc i} properties of the full sample and of the subsample of galaxies with a close companion, by morphological type (col.~1). For each category, we list the number of galaxies $N$, the mean total H{\sc i} mass, and the mean H{\sc i} mass normalised to $R$-band luminosity, which can be interpreted as a kind of ``equivalent width'', and is a measure of the reservoir of gas which has not yet been  transformed into stars. The galaxy UGC~6781 has been excluded from the $T=0$ category because of its misclassification (see text). Col.~3 lists the median value for the gas depletion timescale $\tau$ for the whole sample. }
\label{hiprops}
\end{center}
\end{table}

An important property of galaxies in the context of SF is the amount of gas they contain. Ideally, one would like to know the morphological distribution of both the molecular and the atomic components of the gas, but this has not been observed for even a small part of the H$\alpha$GS sample. What is available for all but a handful of galaxies is the total atomic gas mass, which we obtained in the way described in Sect.~2.2. Subdivided by morphological type, the results are shown in Table~\ref{hiprops}. There, we list for each $T$-class, and separately for all galaxies and only for those with a close companion, the total \hi\ mass in units of solar masses, and the \hi\ mass normalised to the $R$-band luminosity, in units of solar mass per solar luminosity. The latter can be compared to the EW as employed in most of the rest of the current study, and can be interpreted as a measure of the unused gas---gas that is in principle available for SF.

We removed the galaxy UGC~6781 (=NGC~3896)  from the $T=0$ class in Table~\ref{hiprops}.
The galaxy has been classified as SB0/a: pec in the RC3 ($T=0$), but this is in fact a misclassification. Buta, Corwin \& Odewahn (2007) reclassify it as either Im pec (Magellanic irregular) or I0 pec (they compare it to NGC~5253 in this category). Buta et al. note a smooth outer disk and a central area of strong recent SF. UGC~6781 has a close companion in the H$\alpha$GS galaxy UGC~6778 (=NGC~3893), and although we do not classify these galaxies as interacting under our criteria, Buta et al. (2007) note that the slight asymmetry in UGC~6778 could be the result of an interaction. UGC~6778 has an \hi\ mass which is anomalously high for a $T=0$ galaxy, which is further evidence for its misclassification in the RC3. 

Several interesting results show up in Table~\ref{hiprops}, which we describe briefly here.

\subsubsection{Total and normalised HI mass vs $T$-type}

The total \hi\ mass is highest in galaxies of intermediate types, $T=4$ and 5. It is lower in galaxies of earlier and later types, with the lowest gas masses measured in the latest-type galaxies, those of type Sm and Im. This has been well known for a long time, and has been reviewed, for instance, by Roberts \& Haynes (1994).

Normalising the \hi\ mass by $R$-band luminosity yields a rather different picture. This parameter is much higher in the latest-type galaxies ($T=9$ and 10, or Sm and Im), and there is now a gradual trend of increasing $M_{\rm HI}/L_R$ with increasing $T$-type (see also Roberts \& Haynes 1994). This confirms that the gas fraction increases with morphological type throughout the Hubble sequence, most probably as a consequence of pronounced SF activity in the history of the galaxies of earlier morphological type, offset by those of late types which have not been able to transform much of their gas reservoir into stars. Further discussion of these results, and of the important consequences for galaxy formation and evolution they imply, is beyond the scope of this paper.

\subsubsection{Gas depletion timescales}

Median gas depletion timescales, $\tau$, as defined in Sect.~2.2, are given for each morphological type in Table~\ref{hiprops}. Because the number of galaxies per type bin which have a close companion is too small, and because $\tau$ varies with type, we could not produce reliable numbers for the sub-sample of galaxies with a close companion. Across the whole sample, the median gas depletion timescales are of order $10^9-10^{10}$\,yr. They are shorter for galaxies of type $T=2-5$ (Sab$-$Sc) and longer for galaxies of earlier and later types: presumably because of lower SFR in the former, and because of higher gas quantities in the latter. If the galaxies of type $T=2-5$ in our sample would continue to form stars at their current rate, and without any gas infall, they would deplete their gas in $4-5\times10^9$\,yr.

\subsubsection{HI and the presence of close companions}

Table~\ref{hiprops} also shows the \hi\ mass and $M_{\rm HI}/L_R$ separately for those galaxies with close companions. Although the numbers of galaxies in many of the $T$ bins are very small, we have chosen not to make larger bins because of the obvious changes in the \hi\ parameters with morphological type. Nonetheless, it is clear that Table~\ref{hiprops} does not provide significant evidence for different \hi\ mass, whether total or normalised to $L_R$, in those galaxies with close companions compared to the overall population. This can be compared with the results of Bushouse (1987), who found that the median atomic gas content of interacting galaxies is not different from that of isolated galaxies, and also that \hi\ content and SFR do not correlate. 

We note a subtle effect in the following sense, through. Most of the \hi\ masses are comparable between the galaxies with and without a close companion, although some in the latter category and of intermediate Hubble types show a slight enhancement. The values for $M_{\rm HI}/L_R$ are lower for the galaxies with close companions in all but three of the morphological type bins, though, which is presumably due to the increased $R$-band luminosity across that category, as noted in Sect.~4.1 in relation to the EW measurements. The differences are not statistically significant, however, and we conclude that the presence or absence of a close companion has no relation to the total or normalised \hi\ mass in galaxies. 


\section{Interacting and merging galaxies in the sample}\label{interactions}

\begin{figure*}
\begin{center}
\includegraphics[width=0.95\textwidth]{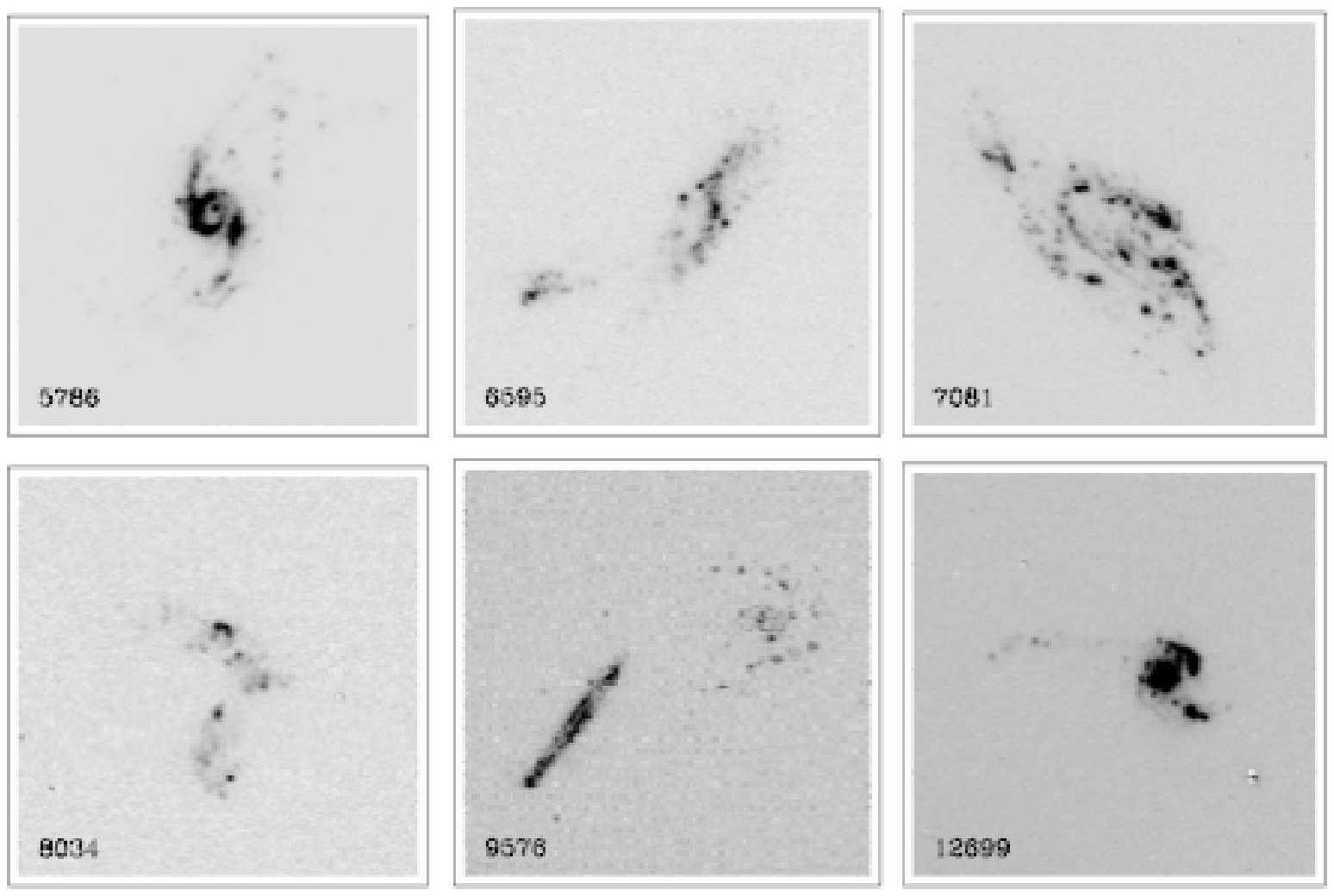}
\end{center}
\caption{\ha GS \ha\ continuum-subtracted images of the interacting galaxies in our sample, and of the advanced merger UGC~5786. Galaxies have been identified by their UGC number. North is up and east to the left in all images. Area shown is 165 (UGC 5786, 6595, 8034, 12699$+$12700), 265 (UGC~7081), and 500\,arcsec (UGC~9576$+$9579) across. For UGC~4541 the quality of the \ha\ image is compromised by the presence of a bright star, so for this galaxy we refer to the broad-band image shown in Fig.~1.}
\label{interactions_figure}
\end{figure*}

Although interacting galaxies and ongoing mergers are too rare in the H$\alpha$GS sample to base our study exclusively on their properties, we can identify a number of such galaxies from the sample of those which have close companions. We thus made a visual check of the images as presented in Fig.~1 to select those galaxies which are interacting or merging. In order to be classified as such, a H$\alpha$GS galaxy must form a pair in which the galaxies must either be overlapping or very close, {\it and} must show signs of either physical contact or tidal disruption as a result of the collision with a companion galaxy. 

A number of close pairs (galaxies separated by around one diameter) where the latter signs were absent have {\it not} been classified as ``interacting'', this category includes the galaxies UGC 858 (whose famous friend is the shell galaxy NGC~474), UGC~7523, UGC~12447, and the M51-lookalike system composed of UGC~6778 and UGC~6781. The galaxy UGC~3740 looks tidally distorted but has not been selected as ``interacting" because the companion is not close enough. In addition, a galaxy such as UGC~5786 (NGC~3310 or Arp 217) is in \Ha GS but can not be included here because it does not qualify as having a close companion under the criteria set out in Section~3 (it is a more advanced merger). The latter two galaxies do have high values for both the SFR (13.5 and 11.2\,$M_{\odot}$\,yr$^{-1}$ for UGC~3740 and 5786, respectively) and the EW (6.1 and 10.9\,nm) (see Section~\ref{topten}). 

\begin{table*}
\begin{center}  
\begin{tabular}{cccccccccccccccc}
\hline
\hline
UGC & NGC & Arp & \multicolumn{2}{c}{Separation} & SFR & EW  & $\tau$ & $\Sigma$ & $\Sigma_{20}$ & $\log\,L_{\rm IR}$ & \multicolumn{5}{c}{Rank}\\
&&&&&&&&&&& SFR & EW & $\Sigma$ & $\Sigma_{20}$ & $\tau$\\
\hline
4541 & 2648 & 89     & 12.9 & 55 & 0.30 & 0.2 & 8.4 & 0.0004 & 0.0045 & -- & 175 & 323 & 311 & 251 & 133\\
6595 & 3769 & 280  & 4.8 & 10 & 0.39 & 2.9 & 10.5 & 0.0033 & 0.023 & -- & 157 & 171 & 172 & 98 & 161\\
7081 & 4088 & 18     & 42 & 40 & 3.06 & 4.4 & 3.3 & 0.010 & 0.027 & 10.25 & 23 & 76 & 42 & 87 & 53\\
8034 & 4809 & 277   & 4 & 3 & 0.51 & 8.0 & 6.6 & -- & -- & -- & 134 & 15 & -- & -- & 111\\
9576 & 5774 & --        & 35 & 110 & 2.7 & 5.2 & 6.6 & 0.0059 & 0.021 & 10.78 & 27 & 49 & 97 & 110 & 112\\
9579 & 5775 & --        & 35 & 110 & 5.0 & 2.5 & 6.7 & 0.0081 & 0.054 & 10.78 & 8 & 193 & 62 & 45 & 114\\
12699 & 7714 & 284 & 18 & 30 & 7.8 & 9.6 & 1.8 & 0.031 & 0.61 & 10.72 & 4 & 9 & 7 & 2 &18\\
12700 & 7715 & 284 & 18 & 30 & 0.0 & 0.0 & $>100$ & -- & -- & 10.72 & 327 & 327 & -- & -- & 327\\
\hline
\end{tabular}
\caption[]{Names (UGC, NGC, Arp; cols. 1, 2, and 3) and properties of the interacting/merging systems among our sample galaxies, in order of increasing UGC number: separation in projected distance (in kpc; col.~4) and velocity (in km\,s$^{-1}$; col.~5); SFR in  $M_{\odot}\,{\rm yr}^{-1}$ and EW in nm (cols.~6 and 7); the gas depletion timescale $\tau$, in $10^9$\,yr (col.~8); the SFR per unit area ($\Sigma$) across the disk out to $r_{24}$, and in the central region out to 0.2\,$r_{24}$ ($\Sigma_{20}$; in units of $M_{\odot}\,{\rm yr}^{-1}\,{\rm kpc}^{-2}$; cols.~9 and 10);
and the log of the total IR luminosity in solar units (from Sanders et al. 2003; col.~11; note that values are given for the combined systems UGC~9576+9579 and UGC~12699+12700).  Where no values can be calculated a field is indicated by `--'. We also list the rank number of the galaxy in the lists of all 327 sample galaxies as sorted on SFR, \ha\ EW, $\Sigma$, $\Sigma_{20}$, and $\tau$ (columns~12--16). Data on SFR and EW from Paper~I.}
\label{interactions_table}
\end{center}
\end{table*}

The strict set of criteria we employed reduces the list of interacting/merging systems among the H$\alpha$GS sample to the 6 systems (8 H$\alpha$GS galaxies) listed in Table~\ref{interactions_table}, but does ensure that these galaxies are without reasonable doubt interacting or merging. \ha\ images of these galaxies, plus of the advanced merger UGC~5786, are shown in Fig.~\ref{interactions_figure}. All systems but one  have been classified by Arp (1966) as interacting systems, and their Arp numbers are given in Table~\ref{interactions_table}. 
We can use our H$\alpha$GS sample to make a very rough estimate of how many galaxies in the local Universe are interacting (in Sect.~3 we already saw that just under 15\% of the sample galaxies have a close companion). Using all of H$\alpha$GS, we thus find that only 2\% of nearby galaxies are interacting. Even ignoring the 117 galaxies in H$\alpha$GS that are fainter than $M_R=-17.5$\,mag, and counting UGC~3740 and 5786 (see above) as interacting, the fraction is still only 4\%. So interacting galaxies are indeed rare.

\subsection{LIRGs and ULIRGs}\label{lirgs}

There are no ultra-luminous infrared galaxies (ULIRGs) in H$\alpha$GS, confirming their rarity in the local Universe. Only one luminous infrared galaxy (LIRG) is included in H$\alpha$GS, namely UGC~3429 or NGC~2146. The galaxy has a total infrared (IR) luminosity of $\log\,L_{\rm IR}=11.07\,L_\odot$ (Sanders et al. 2003), does not have a close companion, and has a SFR of 2.7\,$M_\odot\,{\rm yr}^{-1}$, an EW of 2.5\,nm, and a gas depletion timescale $\tau=0.86\times10^9$\,yr (H$\alpha$GS values). This galaxy has been widely studied, but we just mention the rather recent paper by Greve et al. (2006), in which the interested reader can find a comprehensive list of other work on this LIRG. 

Table~\ref{interactions} lists the total IR luminosities of those galaxies for which a value is given by Sanders et al. (2003). For the pairs UGC~9576$+$9579 and UGC~12699$+$12700 the large IRAS beam yields only one measurement for each pair. UGC~5786, in H$\alpha$GS but not in Table~\ref{interactions} (see above) has a $\log\,L_{\rm IR}=10.61\,L_\odot$. 

The numbers presented here illustrate how rare IR-luminous galaxies are in the local Universe. Sanders et al. (2003) list only 28 LIRGs ($11<\log\,L_{\rm IR}<12$) with $v_{\rm sys}<4000$\,km\,s$^{-1}$, and of these, 19 have $\log\,L_{\rm IR}\leq 11.1$, and only two (NGC~3256 and NGC~3690) have $\log\,L_{\rm IR} > 11.5$ (only two ULIRGs have $v_{\rm sys}<10000$\,km\,s$^{-1}$, UGC~9913 being the closest at $v_{\rm sys}=5450$\,km\,s$^{-1}$). So although these may represent luminous starbursts in the best definable way (see Sect.~7), their role in at least the local Universe is limited purely because of their rarity. 


\section{Characteristics of H$\alpha$GS galaxies with enhanced  star formation}\label{topten}

In this Section we discuss the characteristics of those H$\alpha$GS galaxies that have, by one or other measure, enhanced SF. To illustrate the properties of these galaxies as defined under different sets of parameters, we therefore select the top ten of H$\alpha$GS galaxies which have the highest overall SFR, EW, and SFR per unit area ($\Sigma$), as well as the ten galaxies with the highest central $\Sigma$, measured over the central 20\% of $r_{24}$ only (where $r_{24}$ is the radius corresponding to the $R=24$ isophote), and the ten galaxies with the lowest gas depletion timescale. The results presented here will be used in the next Section (7.1) to discuss critically the possible definition of the term ``starburst'', where we will show that no reasonable definition leads to a well-defined group of objects.

\subsection{Top ten SFR galaxies}

\begin{table*}
\begin{center}  
\begin{tabular}{cccccccccccccccc}
\hline
\hline
UGC & Other names & Type & SFR & EW & $\tau$ & $\Sigma$ & $\Sigma_{20}$ & $D$ & $M$ & Co. & Int. & \multicolumn{4}{c}{Rank}\\
&&&&&&&&&&&& EW & $\Sigma$ & $\Sigma_{20}$ & $\tau$ \\
\hline
3740 & NGC~2276, Arp 25 & SABc & 13.5 & 6.1 & 1.3 & 0.022 & 0.089 & 33.2 & $-$20.7 & -- & -- & 31 & 13 & 25 & 11\\
5786 & NGC~3310, Arp~217 & SABbc & 11.2 & 10.9 & 1.2 & 0.060 & 1.010 & 18.2 & $-$20.2 & -- & -- & 8 & 1 & 1 & 10\\
2855 & & SABc & 9.8 & 3.4 & 1.7 & 0.032 & 0.11 & 17.5 & $-$17.7 & -- & -- & 129 & 4 & 14 & 17\\
12699 & NGC~7714, Arp~284 & SBb & 7.8 & 9.6 & 1.8 & 0.031 & 0.61 & 31.0 & $-$19.5 & Y & Y & 9 & 7 & 2 & 18\\
12343 & NGC 7479 & SBc & 6.6 & 3.5 & 2.9 & 0.010 & 0.040 & 26.9 & $-$20.6 & -- & -- & 121 & 39 & 62 & 41\\
9926 & NGC 5962 & SAc & 6.4 & 3.3 & 1.8 & 0.012 & 0.070 & 31.2 & $-$20.5 & -- & -- & 134 & 29 & 32 & 20\\
6778 & NGC 3893 & SABc & 5.6 & 5.0 & 2.9 & 0.013 & 0.075 & 18.5 & $-$20.2 & Y & -- & 56 & 24 & 29 & 39\\
9579 & NGC~5774 & SBc & 5.0 & 2.5 & 6.7 & 0.0081 & 0.054 & 28.8 & $-$20.2 & Y & Y & 193 & 62 & 45 & 114\\
12447 & NGC 7541 & SBbc & 4.4 & 3.4 & 5.0 & 0.0028 & 0.018 & 29.9 & $-$20.0 & Y & -- & 130 & 192 & 121 & 85\\
11604 & NGC 6951 & SABbc & 4.3 & 1.6 & 2.1 & 0.010 & 0.066 & 20.2 & $-$19.9 & -- & -- & 258 & 41 & 33 & 24\\
\hline
\end{tabular}
\caption[]{Ranked list of the top ten galaxies from the H$\alpha$GS sample in terms of SFR. We list UGC, NGC, and Arp numbers (columns~1, 2); morphological type from the RC3 (column~3); SFR (in units of $M_{\odot}\,{\rm yr}^{-1}$), \ha\ EW (nm), gas depletion time ($10^9$\,yr), SFR per unit area ($\Sigma$) across the disk out to $r_{24}$, and in the central region out to 0.2\,$r_{24}$ ($\Sigma_{20}$; latter two in units of $M_{\odot}\,{\rm yr}^{-1}\,{\rm kpc}^{-2}$; columns~4--8); distance in Mpc (column~9); absolute magnitude (calculated from NED $B_T$ magnitudes converted to absolute magnitudes using the distance $D$; column~10); whether the galaxy has a close companion or not (following Sect.~\ref{definitions}; column 11) and is interacting or not (Section~\ref{interactions}; column~12); the rank number of the galaxy in the lists of all 327 sample galaxies as sorted on \ha\ EW, $\Sigma$, $\Sigma_{20}$, and $\tau$ (columns~13--16). Data on SFR, EW, distance, and absolute magnitude from Paper~I. Comments for individual galaxies:\\
UGC~3740: tidally distorted\\
UGC~5786: advanced merger\\
UGC~2855: possibly a pair with UGC~2866\\
UGC~12699: LINER, interacting, see Sect.~5\\
UGC~12343: LINER/Sy\\
UGC~9579: interacting, see Sect.~5\\
UGC~12447: close pair, see Sect.~5\\
UGC~11604: Sy, circumnuclear ring.}
\label{top10SFR}
\end{center}
\end{table*}

To understand better what the characteristics are of those galaxies that show particularly high values of their SFR, we list in Table~\ref{top10SFR} the top ten H$\alpha$GS galaxies with the highest SFR. From the top ten in SFR, we can make the following observations:

\begin{itemize}

\item All galaxies except one also have NGC numbers, three have been listed as Arp objects.

\item All galaxies are of morphological type 3--5, and nine of the ten are of types 4 (bc) or 5 (c).

\item Two of the ten galaxies have an \ha\ EW below the median in the sample (which is 3.0\,nm; the average EW is 3.5\,nm), several others are not far off the median.

\item Gas depletion timescales $\tau$ are rather short, and below 1\,Gyr for all except UGC~9579 and UGC~12447. For comparison, the median value for $\tau$ across the H$\alpha$GS sample is $1.1\times10^{10}$\,yr, and across all Sc-type galaxies it is $4.0\times10^{9}$\,yr.

\item All galaxies are more distant and brighter than the mean or median value in H$\alpha$GS (mean values for $D$ and $M$ are 17.3\,Mpc and $-17.3$\,mag, respectively, median values are 17.3\,Mpc and $-17.6$\,mag).

\item Galaxies with close companions, interacting, or showing morphological evidence for advanced merging activity are common in the top ten.

\end{itemize}

We thus find predominantly large and bright spirals, a significant fraction of which are interacting or merging.

\subsection{Top ten EW galaxies}

\begin{table*}
\begin{center}  
\begin{tabular}{cccccccccccccccc}
\hline
\hline
UGC & Other names & Type & SFR & EW & $\tau$ & $\Sigma$ & $\Sigma_{20}$ & $D$ & $M$ & Co. & Int. & \multicolumn{4}{c}{Rank}\\
&&&&&&&&&&&& SFR & $\Sigma$ & $\Sigma_{20}$ & $\tau$\\
\hline
3847 & NGC 2363 & Im & 0.02 & 33.9 & -- & 0.041 & 0.24 & 2.9 & $-$11.8 & -- & -- & 220 & 2 & 4 & --\\
3851 & NGC 2366 & IBm & 0.2 & 23.6 & 5.1 & 0.033 & 0.017 & 2.9 & $-$15.9 & -- & -- & 185 & 3 & 126 & 86\\
2302 & & SBm & 0.3 & 20.5 & 19.1 & 0.0042 & 0.0023 & 12.8 & $-$16.4 & -- & -- & 172 & 140 & 280 & 227\\
8098 & NGC 4861, Arp 266 & SBm & 1.1 & 18.2 & 2.7 & 0.0020 & 0.0073 & 11.0 & $-$17.3 & -- & -- & 73 & 239 & 209 & 33\\
6833 & NGC 3930 & SABc & 1.6 & 13.5 & 3.8 & 0.011 & 0.022 & 17.6 & $-$18.1 & -- & -- & 48 & 33 & 100 & 64\\
5829 & & Im & 0.3 & 12.3 & 9.1 & 0.0065 & 0.029 & 8.6 & $-$15.9 & -- & -- & 177 & 91 & 85 & 140\\
5786 & NGC~3310, Arp~217 & SABbc & 11.2 & 10.9 & 1.2 & 0.060 & 1.010 & 18.2 & $-$20.2 & -- & -- & 2 & 1 & 1 & 10\\
12699 & NGC~7714, Arp~284 & SBb & 7.8 & 9.6 & 1.8 & 0.031 & 0.61 & 31.0 & $-$19.5 & Y & Y & 4 & 7 & 2 & 18\\
9018 & NGC 5477 & SAm & 0.03 & 9.2 & 23.0 & 0.010 & 0.063 & 4.3 & $-$13.8 & -- & -- & 271 & 43 & 38 & 246\\
12732 & & Sm & 0.12 & 8.8 & 27.5 & 0.0033 & 0.0033 & 8.9 & $-$ 16.0 & -- & -- & 212 & 176 & 270 & 260 \\
\hline
\end{tabular}
\caption[]{Ranked list of the top ten galaxies in H$\alpha$GS in terms of \ha\ EW. Columns as in Table~\ref{top10SFR}, except column~13 which now shows the rank number in the list as sorted on SFR. Comments for individual galaxies (not repeated from Table~\ref{top10SFR}):\\
UGC~3847: H{\sc ii} complex in UGC~3851\\
UGC~8098: BCD galaxy\\
UGC~9018: M101 satellite.}
\label{top10EW}
\end{center}
\end{table*}

In contrast, we can make rather different observations for our top ten EW galaxies. We note that UGC~3847, although catalogued individually as a galaxy, is in fact an H{\sc ii} region on the outskirts of UGC~3851. As a combined system, they top the EW ranking list with a combined SFR of 0.22\,$M_{\odot}\,{\rm yr}^{-1}$ and EW of 24.2\,nm (both numbers dominated by UGC~3851 which is some 3\,mag brighter). Some observations are:

\begin{itemize}

\item Three galaxies have been listed as Arp objects, and two of those (UGC~5786 and UGC~11699) are also in the top ten SFR list. Three galaxies do not have an NGC designation.

\item All but three of the galaxies are of the very latest morphological types ($T=9, 10$), and two of the remaining three are UGC~5786 (advanced merger) and UGC~11699 (interacting).

\item Only four of the ten galaxies have a SFR larger than the mean and even the median SFR in H$\alpha$GS (which are 0.9 and 0.3\,$M_{\odot}\,{\rm yr}^{-1}$, respectively). All except one (the well-known starburst and blue compact dwarf [BCD] UGC~8098, a.k.a. NGC~4861 or Arp~266) of the late-type galaxies (9 or 10) are below the mean and median SFR.

\item Gas depletion timescales $\tau$ are rather long for most galaxies. The median value for $\tau$ across the H$\alpha$GS sample is $1.1\times10^{10}$\,yr, and across all m-type galaxies ($T=8-10$) it is $1.9\times10^{10}$\,yr.

\item The seven late-type galaxies are all closer than the mean/median distance in H$\alpha$GS, and are also all fainter in absolute magnitude than the mean/median across the whole sample. Only three of the ten galaxies are more distant, and brighter.

\item Only the two galaxies that are also in the SFR top ten are interacting or merging. None of the other galaxies has a close companion (cf. Section~\ref{definitions}) although a few may be in groups (but we cannot judge whether this would be more often the case for these galaxies than across the whole H$\alpha$GS sample).

\end{itemize}

Galaxies with high \ha\ EW are thus predominantly galaxies with very small amounts of $R$-band continuum emission rather than very large amounts of \ha\ emission, and thus predominantly star-forming late-type dwarf or irregular galaxies.

\subsection{Top ten SFR per unit area galaxies}

\begin{table*}
\begin{center}  
\begin{tabular}{cccccccccccccccc}
\hline
\hline
UGC & Other names & Type & SFR & EW & $\tau$ & $\Sigma$ & $\Sigma_{20}$ & $D$ & $M$ & Co. & Int. & \multicolumn{4}{c}{Rank}\\
&&&&&&&&&&&& SFR & EW & $\Sigma_{20}$  & $\tau$\\
\hline
5786 & NGC~3310, Arp~217 & SABbc & 11.2 & 10.9 & 1.2 & 0.060 & 1.010 & 18.2 & $-$20.2 & -- & -- & 2 & 8 & 1 & 10\\
3847 & NGC 2363 & Im & 0.02 & 33.9 & -- & 0.041 & 0.24 & 2.9 & $-$11.8 & -- & -- & 220 & 1 & 4 & --\\
3851 & NGC 2366 & IBm & 0.2 & 23.6 & 5.1 & 0.033 & 0.017 & 2.9 & $-$15.9 & -- & -- & 185 & 2 & 126 & 86\\
2855 & & SABc & 9.8 & 3.4 & 1.7 & 0.032 & 0.11 & 17.5 & $-$17.7 & Y & -- & 3 & 129 & 14 & 17\\
5398 & NGC~3077 & Im & 0.08 & 4.2 & 17.6 & 0.032 & 0.54 & 2.1 & $-$16.0 & -- & -- & 239 & 87 & 3 & 218\\
2455 & NGC~1156 & IBm & 0.28 & 7.4 & 3.5 & 0.032 & 0.16 & 4.9 & $-$16.1 & -- & -- & 180 & 17 & 7 & 60\\
12699 & NGC~7714, Arp~284 & SBb & 7.8 & 9.6 & 1.8 & 0.031 & 0.61 & 31.0 & $-$19.5 & Y & Y & 4 & 9 & 2 & 18\\
7414 & NGC~4299 & SABd & 0.03 & 7.6 & 2.1 & 0.026 & 0.14 & 2.4 & $-$14.0 & Y & -- & 266 & 16 & 10 & 25\\
3711 & NGC~2337 & IBm & 0.23 & 5.1 & 6.0 & 0.025 & 0.047 & 7.6 & $-$16.5 & -- & -- & 183 & 51 & 51 & 103\\
6869 & NGC~3949 & SAbc & 2.4 & 5.3 & 2.2 & 0.024 & 0.10 & 13.9 & $-$19.2 & -- & -- & 30 & 46 & 18 & 27\\
\hline
\end{tabular}
\caption[]{Ranked list of the top ten galaxies in H$\alpha$GS in terms of SFR per unit area ($\Sigma$). Columns as in Table~\ref{top10EW}, except column~14 which now shows the rank number in the list as sorted on EW. Comments for individual galaxies:\\
UGC~5398; in M81 group.}
\label{top10Sigma}
\end{center}
\end{table*}

A variation on using the SFR to measure the SF activity of a galaxy is to normalise the SFR to the size of the galaxy, or its area. Such a distinction may help separate large galaxies that form stars throughout (most of) their disk (such as UGC~12343=NGC~7479 in Table~\ref{top10SFR}) from smaller galaxies where the SFR per unit area is much higher. We have calculated this for the H$\alpha$GS galaxies using the radius $r_{24}$ for the normalization, and show the ten galaxies with the highest values of SFR per unit area ($\Sigma$) in Table~\ref{top10Sigma}. We can observe the following characteristics:
 
 \begin{itemize}
 
 \item Only three of the top ten $\Sigma$ galaxies are also in the top ten SFR list (Table~\ref{top10SFR}), and most have SFRs below the median across H$\alpha$GS. The latter obviously have large $\Sigma$ values because of their small size, and not because of their large SFRs.
 
 \item Most galaxies have also modest EW values, although in all cases above the median EW across H$\alpha$GS. So $\Sigma$ correlates better with EW than with SFR.
 
 \item Most galaxies in the top ten $\Sigma$ list are of late morphological type, one $T=8$ and five $T=10$, and these galaxies are all nearby and of very modest absolute magnitude.
 
 \item Two of our by now familiar merger/interaction galaxies have large $\Sigma$ values as well as the high SFR and EW that we already noted before. 
 
 \item Gas depletion timescales are not particularly noteworthy, although in all but one cases shorter than the median value across H$\alpha$GS. 
 
 \end{itemize}
  
\subsection{Top ten central SFR per unit area galaxies}

\begin{table*}
\begin{center}  
\begin{tabular}{cccccccccccccccc}
\hline
\hline
UGC & Other names & Type & SFR & EW & $\tau$ & $\Sigma$ & $\Sigma_{20}$ & $D$ & $M$ & Co. & Int. & \multicolumn{4}{c}{Rank}\\
&&&&&&&&&&&& SFR & EW & $\Sigma$ & $\tau$\\
\hline
5786 & NGC~3310, Arp~217 & SABbc & 11.2 & 10.9 & 1.2 & 0.060 & 1.010 & 18.2 & $-$20.2 & -- & -- & 2 & 8 & 1 & 10\\
12699 & NGC~7714, Arp~284 & SBb & 7.8 & 9.6 & 1.8 & 0.031 & 0.61 & 31.0 & $-$19.5 & Y & Y & 4 & 9 & 7 & 18\\
5398 & NGC~3077 & Im & 0.08 & 4.2 & 17.6 & 0.032 & 0.54 & 2.1 & $-$16.0 & -- & -- & 239 & 87 & 5 & 218\\
3847 & NGC 2363 & Im & 0.02 & 33.9 & -- & 0.041 & 0.24 & 2.9 & $-$11.8 & -- & -- & 220 & 1 & 2 & --\\
10470 & NGC~6217, Arp~185 & SBbc & 2.9 & 3.8 & 5.7 & 0.020 & 0.21 & 21.2 & $-$19.8 & -- & -- & 24 & 106 & 16 & 98\\
2183 & NGC~1056, Mrk~1183 & Sa & 0.8 & 2.3 & 10.8 & 0.009 & 0.16 & 18.6 & $-$18.0 & -- & -- & 93 & 215 & 52 & 163\\
2455 & NGC~1156 & IBm & 0.28 & 7.4 & 3.5 & 0.032 & 0.16 & 4.9 & $-$16.1 & -- & -- & 180 & 17 & 6 & 60\\
3429 & NGC~2146 & SBab & 2.7 & 2.5 & 3.3 & 0.007 & 0.14 & 14.5 & $-$19.4 & -- & -- & 28 & 195 & 70 & 52\\
7096 & NGC~4102 & SABb & 1.7 & 2.6 & 1.1 & 0.012 & 0.14 & 15.2 & $-$18.9 & -- & -- & 41 & 191 & 31 & 8\\
7414 & NGC~4299 & SABd & 0.03 & 7.6 & 2.1 & 0.026 & 0.14 & 2.4 & $-$14.0 & Y & -- & 266 & 16 & 8 & 25\\
\hline
\end{tabular}
\caption[]{Ranked list of the top ten galaxies in H$\alpha$GS in terms of central SFR per unit area ($\Sigma$$_{20}$, measured over the inner 20\% of $r_{24}$ only). Columns as in Table~\ref{top10Sigma}, except column~15 which now shows the rank number in the list as sorted on $\Sigma$. Comments for individual galaxies:\\
UGC~10470: Sy 2\\
UGC~2183: Sy 2\\
UGC~3429: LIRG, see Sect.~5\\
UGC~7096: LINER.}
\label{top10Sigma20}
\end{center}
\end{table*}

A variation on the use of $\Sigma$, the SFR per unit area across a galaxy, is to use $\Sigma_{20}$, which we defined as the central SFR per unit area, or $\Sigma$ determined only for the area enclosed in the inner 20\% of $r_{24}$. The ten galaxies with highest $\Sigma_{20}$ values are listed in Table~\ref{top10Sigma20}. We make the following comments:

\begin{itemize}

\item Four of these ten galaxies have SFRs below the median across H$\alpha$GS, three have below-median EWs, and none has below-median $\Sigma$ values.

\item The majority (6/10) of top ten $\Sigma_{20}$ galaxies are also in the top ten $\Sigma$. The $\Sigma$ and $\Sigma_{20}$ parameters are vaguely related, and $\Sigma_{20}$ appears to possess less discriminating power in identifying powerful SF in galaxies than might have been expected.

\item In comparison with the $\Sigma$ top ten list, fewer galaxies of late morphological type, and more with evidence for nuclear activity appear in the $\Sigma_{20}$ list (the latter not surprisingly).

\item The single IR-luminous galaxy in H$\alpha$GS, the LIRG UGC~3429, makes its one and only appearance in the top ten lists in the $\Sigma_{20}$ table. In all other parameters considered it is entirely unremarkable, and in terms of EW in fact even below the median of the whole H$\alpha$GS sample.

\end{itemize}

\subsection{Ten galaxies with the lowest gas depletion timescale}

\begin{table*}
\begin{center}  
\begin{tabular}{cccccccccccccccc}
\hline
\hline
UGC & Other names & Type & SFR & EW & $\tau$ & $\Sigma$ & $\Sigma_{20}$ & $D$ & $M$ & Co. & Int. & \multicolumn{4}{c}{Rank}\\
&&&&&&&&&&&& SFR & EW & $\Sigma$ & $\Sigma_{20}$\\
\hline
4645 & NGC 2681 & SAB0/a & 0.57 & 0.9 & 0.065 & 0.098 & 0.0057 & 12.3 & $-$19.4 & -- & -- & 128 & 295 & 104 & 35\\
7328 & NGC~4245 & SB0/a & 0.20 & 0.9 & 0.25 & 0.0039 & 0.051 & 10.4 & $-$17.8 &  -- & -- & 191 & 296 & 152 & 48\\
7405 & NGC~4293 & SB0/a & 0.78 & 0.7 & 0.27 & 0.0015 & 0.015 & 17.6 & $-$20.0 &  -- & -- & 95 & 307 & 263 & 138\\
4375 & -- & SABc & 1.23 & 1.6 & 0.29 & 0.0028 & 0.0056 & 30.9 & $-$20.7 & Y & -- & 67 & 261 & 197 & 239\\
7054 & NGC~4064 & SBa & 0.69 & 1.3 & 0.54 & 0.0019 & 0.032 & 17.6 & $-$19.0 &  -- & -- & 107 & 283 & 242 & 73\\
7315 & NGC~4237 & SABbc & 0.86 & 1.8 & 1.0 & 0.0065 & 0.047 & 17.6 & $-$18.7 &  -- & -- & 91 & 251 & 89 & 52\\
6272 & NGC~3593 & SA0/a & 0.31 & 2.6 & 1.1 & 0.0057 & 0.092 & 7.1 & $-$17.4 &  -- & -- & 174 & 189 & 102 & 21\\
7096 & NGC~4102 & SABb & 1.7 & 2.6 & 1.1 & 0.012 & 0.14 & 15.2 & $-$18.9 & -- & -- & 41 & 191 & 31 & 8\\
2045 & NGC~972 & Sab & 2.2 & 2.8 & 1.1 & 0.0074 & 0.078 & 18.5 & $-$19.1 &  -- & -- & 33 & 176 & 71 & 27\\
5786 & NGC~3310, Arp~217 & SABbc & 11.2 & 10.9 & 1.2 & 0.060 & 1.010 & 18.2 & $-$20.2 & -- & -- & 2 & 8 & 1 & 10\\
\hline
\end{tabular}
\caption[]{Ranked list of the ten galaxies in H$\alpha$GS with the lowest gas depletion timescale $\tau$. Columns as in Table~\ref{top10Sigma20}, except column~16 which now shows the rank number in the list as sorted on $\Sigma_{20}$. Comments for individual galaxies:\\
UGC~4645: LINER\\
UGC~7328: possibly pair with NGC~4253 at 16.5 arcmin\\
UGC~7405: LINER\\
UGC~6272: Sy2 (NED)
}
\label{top10tau}
\end{center}
\end{table*}

Our final top ten list (or rather bottom ten, in this case; Table~\ref{top10tau}) lists the ten H$\alpha$GS galaxies with the shortest gas depletion timescales, as defined in Sect.~2.2. This timescale indicates how many years a galaxy can continue forming stars at its current SFR before it exhausts its total gas content. This is a very rough estimate, ignoring any inflow into the galaxy, and assuming all gas can take part in the SF, but does give a useful indication of the relation between SF and ``fuel''. Variations on this parameter have been proposed in the literature to define starburst galaxies, as discussed in the next Section. Some facts to note from Table~\ref{top10tau} are the following:

\begin{itemize}

\item With one exception (the advanced merger UGC~5786), {\it none} of the ten shortest $\tau$ galaxies has above-median EW values. Two galaxies have below-median SFRs, and most have very modest $\Sigma$ values.

\item Half of the galaxies (and four out of the five in the top half of Table~\ref{top10tau}) are of the earliest morphological types, $T=0$ or 1. None is of type later than Sc ($T=5$). This is correlated with the low EW values in many of these galaxies.

\item For most of the galaxies in this list, the short gas depletion timescales seems to be the result of a low gas mass rather than a high SFR, reflected in the prominent places occupied in the Table by early-type galaxies. The broad dependence of the median value of $\tau$ on morphological type can also be seen in Table~\ref{hiprops}, although the very lowest values are, apparently, concentrated among the galaxies of earliest morphological type.

\end{itemize}

Only one galaxy consistently appears among the top in each category we considered.  This is the advanced merger UGC~5786, discussed already earlier in this paper. Even so, it barely makes it into the top ten tables for EW (8th) and $\Sigma_{20}$ and $\tau$ (10th place in each). Any other galaxy in H$\alpha$GS, including the starburst BCD UGC~8098 (NGC~4861; Arp~266), the LIRG UGC~3429 (NGC~2146) or the interacting Arp system 284 (UGC~12699$+$12700; NGC~7714$+$7715) fail to make it into the top ten for at least one of the categories (in fact, as noted before, in all {\it but} one in the case of the LIRG). 

\begin{figure}
\begin{center}
\includegraphics[width=0.45\textwidth]{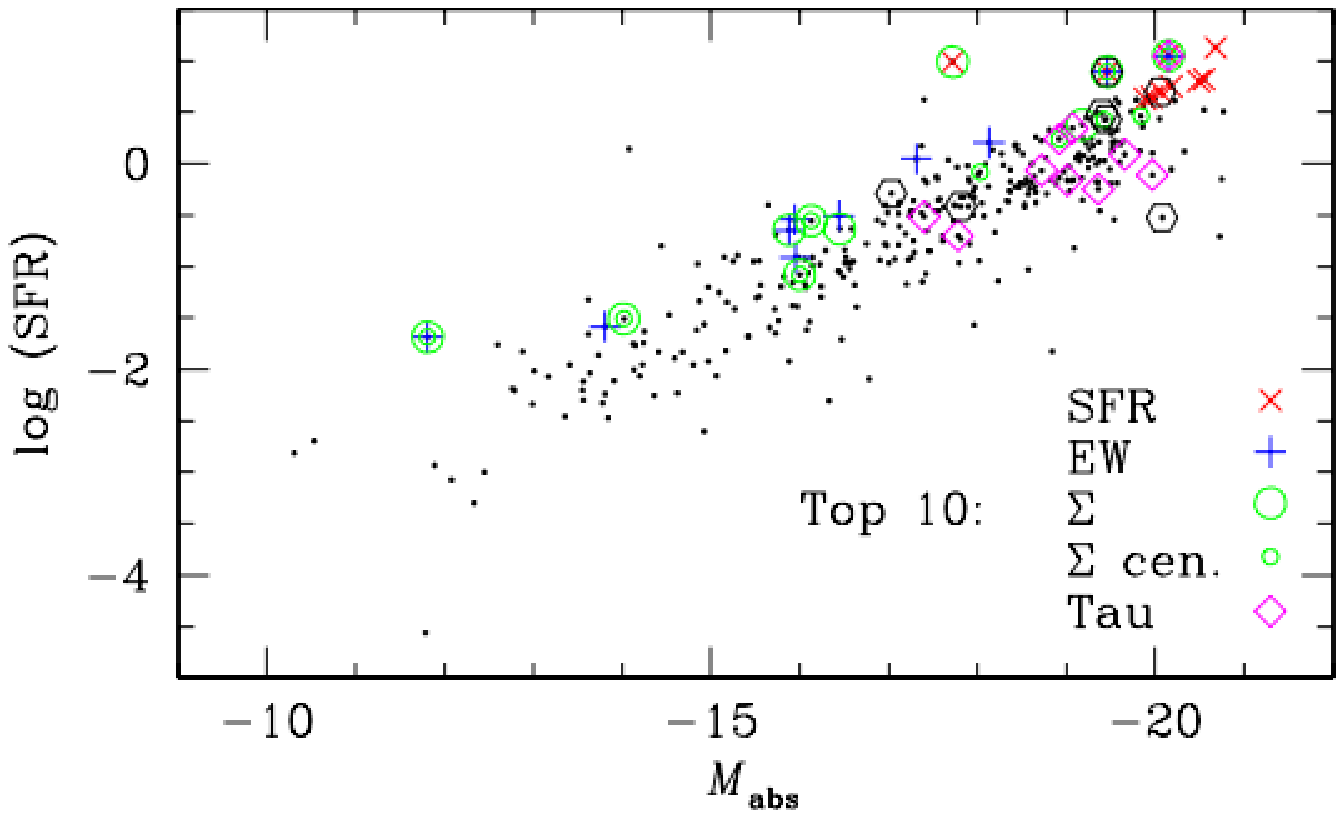}
\end{center}
\caption{As Fig.~\ref{figsfrmabs}, but now indicating the location in the diagram of the galaxies in our various top-tens, as indicated in the legend. In addition, the eight interacting galaxies we defined are identified with black hexagons. Galaxies can appear in more than one category, and some are thus identified by various symbols simultaneously.
}
\label{figtop10}
\end{figure}

Figure~\ref{figtop10} shows where the various top-ten galaxies are located on the diagram of SFR vs absolute magnitude, as shown in Fig.~\ref{figsfrmabs} for all sample galaxies. Obviously, the top-ten SFR galaxies are in the upper right corner of the diagram because that is where the galaxies with largest SFRs are located. The galaxies with the highest EW are, with one exception, located near the upper envelope of the cloud of points, yet they do {\it not} pick out galaxies with the very highest SFRs at their absolute magnitude. In broad agreement with the analogies between these groups of galaxies we noticed before, the galaxies with the highest $\Sigma$ and central $\Sigma$ values are located in an area similar to that of the high EW galaxies. Galaxies with lowest gas depletion times are all in the right side of the diagram (having $M_{abs}<-17$) but by no means trace the upper part of the distribution in SFR. Finally, the eight interacting galaxies, denoted by black open hexagons in Fig.~2, show the spread in SFR discussed before. Some have SFRs among the highest in our sample, whereas others are near or even below the median SFR for their absolute magnitude.

\section{Discussion}\label{discussion}

\subsection{Effects of close companions and interactions on star formation and starbursts}

As shown in Table~\ref{SFR+EW} and described in Section~\ref{results}, across the whole sample of 327 \ha GS galaxies that we analyzed the SFR is significantly enhanced in those galaxies that have a close companion (as defined in Section~3), whereas the EW is not significantly enhanced. Galaxies that are gravitationally interacting or merging are rare, and there are not enough of them in a sample of local galaxies such as ours to allow a statistically meaningful analysis of their properties. But as Table~4 and Sect.~5 show, whereas some interacting/merging galaxies have high SFRs and/or \ha\ EWs, at least as many others are unremarkable in these properties.

\subsubsection{Close companions}

Our main result is that the SFR is enhanced in galaxies which have a close companion, but that the \ha\ EW is not. This implies that the presence of a companion stimulates massive SF in a galaxy, presumably because of tidal effects causing stirring and inflow of the gas in the disk, but that the SFR is increased continuously, rather than in the form of a ``burst". In an instantaneous burst of SF, both the \ha\ line emission and the \ha\ EW drop off rather abruptly after some $10^7$\,yr, roughly the lifetime of the massive O stars that power the \ha\ emission (see, e.g., Leitherer et al. 1999, their Fig. 83). If the SF is continuous, however, as illustrated in Fig.~84 of Leitherer et al., the \ha\ EW remains at a significantly higher level for timescales of $10^8$ and even up to $10^9$\,yr. During this continuous episode of SF, the \ha\ line emission remains at the same, high, level throughout, and the fact that the EW decreases slowly with time is not, as in the case of an instantaneous burst of SF, due to the \ha\ emission decreasing, but rather to the continuum emission slowly being enhanced, due to the slow but steady build-up of an underlying stellar population.

So in the galaxies with companions in our sample the SF must be enhanced but continuously so: enhanced because we measure enhanced \ha\ emission and thus SFR as compared to galaxies without a close companion, and continuously so because the \ha\ EW is much less enhanced. 

The episode of massive SF which appears to have been triggered by the presence of a close companion must have been continuous during a time of order $10^8$ years minimum. This continuously enhanced SFR is not what one would prefer the term star{\it burst} to mean, and we must thus conclude that the presence of a close companion indeed stimulates the formation of massive stars, but does this not in the form of a ``starburst", but continuously over rather long timescales. The reason for this may be that the galaxies with close companions are located in regions of the local Universe where they regularly merge with small gas-rich galaxies, or even gas clouds. These continuous (very) minor mergers may provide the fuel to enhance their SFR, and do this over extended periods of time. The presence of a close companion by our criteria may in that case be a tracer of the presence of gas-rich galaxies or gas clouds in the near vicinity. 

As we noted elsewhere in this paper, it might have been more interesting to study galaxies that are undergoing a proper interaction or merger, rather than the ones we study here, which are honored by the presence of a close companion. But we also noted that interacting galaxies are too rare to perform the kind of statistical study we present here, so let us consider the properties of the sample of close companion galaxies that our \ha GS survey has yielded. Images of all these galaxies are shown in Fig.~1, which shows clearly that this is a mixed bag. Some pairs are of roughly equal luminosity (and thus, one may assume, mass), many other \ha GS galaxies are accompanied by a galaxy that is much brighter or fainter than itself. Some pairs are very close indeed, and we selected some of these as ``interacting", but most are separated by a distance of a few times their diameter. 

Making the selection criteria more restrictive will yield results which lack statistical significance, but we need to take into account that the results we obtain can not be easily compared with others presented in the literature, e.g., those by Barton et al. (2000) and Barton Gillespie et al. (2003). These authors study galaxies at larger distances than ours which are in pairs with in most cases a smaller separation than our \ha GS-selected galaxies. Some of their results, such as an enhanced \ha\ EW, indeed occur mostly among galaxies with separations of at most a few tens of kpc. We reiterate here some of the advantages of using a sample such as \ha GS, as already discussed in the introductory section of this paper, such as the availability of well-resolved \ha\ images for the complete discs of our galaxies; the completeness of our sample, including the important components of dwarf and irregular galaxies which are usually missing from studies of interacting or close-pair galaxies; or the robust selection criteria applied to select our galaxies with close companions, which naturally leads to the availability of a well-matched and comparable control sample. 

Our result of a moderate increase in SFR among galaxies with a close companion is not in contradiction to those reported by Di Matteo et al. (2008), who find from an analysis of two large datasets of numerical simulations that galaxy interactions and mergers at low redshifts trigger only moderate SF enhancements. In the large majority of cases, the SFR is enhanced by less than a factor of five, and only in a few galaxies is the duration of this enhancement larger than 500\,Myr. An important difference is that our study targets not only interacting and merging galaxies but also those with close companions. Nonetheless, it seems clear that major starbursts are not necessarily the result of an interaction, as also reported on the basis of observations by, e.g., Bergvall et al. (2003), Li et al. (2008) and Jogee et al. (2008). 

Our result that the enhancement of the SFR as a result of the presence of a close companion must be due to continuous SF, with a duration of order $10^8$ to $10^9$\,yr, rather than to a burst of instantaneous SF seems, at face value, less easy to reconcile with the findings of Di Matteo et al. (2008), of generally ``short-lived" merger-driven SF episodes. Only in roughly half of their simulated cases does the SF episode last longer than a few times $10^8$\,yr,  and no more than 15\% last longer than $5\times10^8$\,yr. It is probable that interactions and mergers lead to shorter episodes of enhanced SF than the presence of a close companion as investigated statistically by us, and that the galaxies in our sample are not yet in the more advanced interactive stages. On the other hand, the few interacting/merging galaxies in our sample do not show a hugely different behavior. Further detailed study into the statistical properties of local interacting/merging galaxies is clearly needed, but given the paucity of such systems the challenge is significant.

\subsubsection{Interactions and mergers}

We now come back to the interacting galaxies already introduced in the first paragraph of this Section. Only for the purpose of the specific test we are about to describe, we add the advanced merger UGC~5786 to the list of eight interacting galaxies in Table~4. In analogy to what we calculated in Table~1 and discussed in Sect.~4.1, we now find that the mean SFR for those nine interacting galaxies, normalised by the mean for the same morphological type, is $1.90\pm0.67$, whereas the equivalent number for the EW is $1.59\pm0.46$. Although the ratios are larger than unity, indicating enhanced SF activity in the interacting galaxies, this is not a very strong effect (especially as compared to the values listed in Table~1 for the type $3-8$ galaxies), nor is it statistically significant. 

This test confirms that although, as we saw before, some interacting/merging galaxies indeed show high SFR and EW, these few are mostly offset by other interacting galaxies with unremarkable SFRs and EWs. So statistically, even full-blown interactions and mergers do not cause a strong enhancement in the SF activity. Since the shocks and general upheaval in the ISM of merging/interacting galaxies might be expected to lead to enhanced SF, as evidenced in the extreme by the fact that practically all (ULIRGs) or most (LIRGs) of the most extremely star-forming galaxies known are interacting, it is worth considering why interacting galaxies may not have an enhanced SFR. We can identify a number of reasons, including the timescale of the events; the geometry of the merger; and the internal structure of the merger (e.g., the presence/absence of a bulge, or of gas). 

The body of observational evidence that one or several of these processes are involved includes the statistical studies reported above, and the fact that a pair of interacting galaxies sometimes shows enhanced SF in only one of the two galaxies, with the other essentially quiescent (UGC~12700 in Table~4 is a good example). It is hard to pinpoint observationally which mechanism  is at work exactly, but numerical simulations are ideal to shed more light on these issues. Indeed, Mihos \& Hernquist (1996) find that major mergers always lead to strong gaseous inflows, and to ``moderate to intense starburst activity." The structure of the interacting galaxies, more than the orbital geometry of the encounter, is found to influence the level of SF activity, with, e.g., galaxies with dense central bulges reaching higher starburst activity, but at a later stage in the merging than in the case of bulgeless galaxies. The period of enhanced SFR is relatively short, a result also found by Mihos \& Hernquist (1994) in an earlier paper on minor mergers.

Further clues emerge from the statistical study of Di Matteo et al. (2007), who study the SF efficiency in numerical simulations of just over 200 galaxy interactions, with varying galaxy parameters, such as bulge-to-disc ratio and gas mass, and different orbital geometries. The galaxy and orbit parameters are found to affect greatly the outcome of the interaction in terms of the SF history. Di Matteo et al. (2007) conclude that mergers are not in all cases triggers of starburst activity (defined as SFRs ten times higher than those occurring in isolated galaxies), and that, partly because of the short duration of the phase of very high SFR, ``galaxy interactions are not a sufficient condition for converting large quantities of gas mass into new stars". They also find that the enhanced SF occurs mainly in the form of nuclear starbursts, which would be in contradiction to our results on the radial distribution of enhanced SFR and EW (Sect.~4.3) {\it if} the behavior of interactions and mergers and galaxies with close companions were comparable.

We conclude that our general findings on the few interacting/merging galaxies in \ha GS are not contradicted by numerical results in the literature, but also that further observational study of individual cases and especially statistically meaningful samples of interacting and merging galaxies are needed to proceed. The scarcity of such systems in the local Universe, as highlighted elsewhere in this paper, is an obvious impediment to such progress, though.

\subsection{Definitions of starbursts}

Although starbursts are by broad consensus assumed to be important in shaping the Universe as we know it, it is not easy to define an all-encompassing and generally applicable set of criteria to define whether a galaxy, or a zone within it, is a starburst. A number of families of overlapping definitions can be found in the literature, which have been reviewed by, e.g., Gallagher (2005), Heckman (2005), and Kennicutt et al. (2005), and which can be summarised as follows:

\begin{enumerate}

\item {\it Temporarily high SFR}, or a substantially higher SFR at the present day (or during the starburst phase) than on average during the past. Essentially, this is what one measures with the EW as used in this paper, by making the reasonable assumption that the integrated $R$-band emission, used as the denominator in the EW measure, summarizes the past SF activity by tracing the total quantity of evolved stars in a galaxy.

\item {\it Exceptionally high SFR}. In the extreme, this kind of definition will pick up objects like the LIRGs and ULIRGs, or other objects that are detected in surveys mainly or only because of their extreme SFR (the so-called SCUBA-sources may well fall into this category, see Smail et al. 2002). 

\item {\it Unsustainably high SFR}, for instance compared to the amount of available fuel. Heckman (2005) defines a starburst as a galaxy that is forming stars near the maximum possible rate set by causality, a rate so high that all the gas would be consumed within one dynamical time.

\end{enumerate}

The important point is that although each of these three families of definitions has substantial merit and can be defended individually, they are not mutually inclusive. As seen in Section~\ref{topten} and in particular in the Tables~\ref{top10SFR} and \ref{top10EW}, galaxies which would qualify as a starburst under one of the definitions might not even come close under others. 

The galaxies in our sample with the highest EW (definition class 1, above), those with the highest SFRs (class 2) and those with the shortest gas depletion timescales (class 3) are very different indeed. The top ten EW is dominated by small late-type galaxies with insignificant SFRs, the top ten SFR by massive, bright, and relatively distant mid-type spirals, often with evidence for interactions and/or nuclear activity, and the top (bottom, in fact) ten gas depletion timescale by early-type galaxies with extremely little H\,{\sc i} but with a non-zero SFR. The question is whether all or only some of these are ``starbursts"?

At least for the purpose of this paper, we discard the third family of definitions, as exemplified  above by the gas depletion timescale. Because our sample contains galaxies with a wide variation in the gas content, partly but not exclusively a function of morphological type (see Section~4.4.1), among the galaxies with the lowest gas depletion times are many with very little gas in the first place, but with some (usually also very little compared across our sample) massive SF. This has nothing to do with starbursts in any sense of the word, even when only considering galaxies of a narrow range of morphological types. Of course galaxies with high SFRs may also have low gas depletion timescales, but the point here is that whereas a low $\tau$ is often a property of a starburst, it cannot be used to {\it define} it.

The two remaining families of definitions each have their merit, and we propose here to use them side by side, noting that they do mean different things. The first category, in which the SFR is now higher than it has, on average, been in the past of a galaxy or region, can be represented by the EW. Lee et al. (2008) use this definition to define global starbursts in dwarf galaxies in the local Universe, selecting as such those galaxies which, within their broad morphological type class, have an EW more than 3\,$\sigma$ above the median of the log(EW) of the distribution. For the later types, their limit corresponds to an EW of some 10\,nm, and under this definition most of the galaxies in our top ten EW (Table~\ref{top10EW}) would thus be classed as starburst. This would be a literal interpretation of the term, since these galaxies indeed undergo bursts of SF, with much higher SFR than they have had in the past (as judged by their $R$-band continuum light). Note, again, the low absolute SFRs and long gas depletion timescales of most of these galaxies, with the exceptions (with values $\tau<1$\,Gyr and SFR$>1\,M_{\odot}\,{\rm yr}^{-1}$) being the well-known BCD UGC~8098, and two systems with evidence for merging/interaction. 

Most of the ``starbursts" as defined by a high EW are m-type galaxies with very faint absolute magnitude, low mass, and low SFR. The energies involved and integrated masses of stars formed in bursts in such galaxies will thus be limited, although Garnett (2002) shows that they may play an important role in the enrichment of the IGM because small galaxies lose most of their supernova-driven outflowing gas, where massive galaxies will retain it.  Also, as we showed in Paper~IV, galaxies of these very late types (as well as those of the earliest types, some of which have the lowest gas depletion timescales as we saw above) make only a very small contribution to the overall current SFR density in the local Universe, which is completely dominated by bright galaxies of type $3-5$. 

The second category, then, is that of galaxies which simply have high current massive SFRs. These are listed in Table~\ref{top10SFR} and, as discussed above,  include three well-known Arp systems with evidence of morphological features related to recent or ongoing interaction/merger activity. Most of the other galaxies in this top ten are large Sc-type spirals. Two of the Arp systems have  EWs high enough to make it also into our top ten EW. The remaining top ten SFR galaxies have unremarkable EW values, in most cases near or even below the median value for the H$\alpha$GS sample. The gas depletion timescales for all but two of the top ten SFR galaxies are smaller than the median value across H$\alpha$GS, but they are still rather unremarkable, and only that of the advanced merger UGC~5786 (as noted above, the only galaxy to make it into all of our ten top lists)  is in the list of ten galaxies with the shortest $\tau$ values (albeit at number 10). 

So are these high-SFR galaxies ``starbursts"? We do not think so. Some are, and in particular the three Arp systems may qualify (high SFR and EW, low $\tau$). But others, such as NGC~7479, are ``merely" well-known big, bright, star-forming, spiral galaxies. 

We now consider the alternative SFR criterion which we denote as $\Sigma$: the SFR per unit area, and a variation, $\Sigma_{20}$, which is the central $\Sigma$ as measured over the central 20\% of the 24-mag radius of the galaxy. The top ten galaxies in these categories are listed in Tables~\ref{top10Sigma} and \ref{top10Sigma20}. As we saw in Sections~6.3 and 6.4, most of the galaxies which score highly on $\Sigma$ or $\Sigma_{20}$ have in fact unremarkable and in many cases even below-median SFR and EW values, and many are faint and nearby. Whereas one justification for introducing the $\Sigma$ measures might have been an expectation that they pick up galaxies that form very large numbers of stars across limited spatial areas, in fact small galaxies generally turn out to have the highest values.

Exceptionally large IR luminosity is often used to select those rare galaxies in which copious amounts of SF (in some cases accompanied by non-stellar activity) are hidden behind large quantities of dust, which is heated by the illumination from the SF. We will not comment in detail on the merit of using IR-based definitions to define starbursts (though noting that the IR emission will depend on the properties of the ISM close to the sites of SF), but the H$\alpha$GS study does allow us to make two observations. The first is an obvious and well-known one, which is that extreme IR emitters (LIRGs and especially ULIRGs) are extremely rare in the local Universe. Using such sources to trace the SF patterns in the early Universe, while attractive because the sources may be observed with relative ease, implies the caveat that one will only be studying the absolute tip of the iceberg, the very extremes of the overall population of star-forming galaxies.

The second comment is that our study illustrates how badly correlated the IR luminosity and the optical SF parameters can be in the case of dust obscuration. The one LIRG in our H$\alpha$GS sample (UGC~3429=NGC~2146) is an example to note here: it is unremarkable in most of its optical SF properties, and only makes it into one of our top ten lists (the $\Sigma_{20}$ one). Its SF is indeed very heavily obscured.

We must draw the rather unsatisfactory overall conclusion from this discussion that it is very difficult indeed to come up with a comprehensive, physically meaningful, and objective definition of a starburst galaxy. We reject definitions based on gas depletion timescales or comparable parameters, because they will be biased towards gas-poor systems which have nothing to do with starbursts. Definitions based on EW or other parameters which compare current with average past SFRs are sensitive to galaxies that ``burst" in the literal sense of the word, but will be biased towards galaxies with very small past SFRs---which tend to be small galaxies whose starbursts will make little impact beyond the galaxy. Finally, we have seen that a large SFR does not exclusively select starburst systems, but also galaxies that are simply large, luminous, and bright, and that normalising the SFR to area favors small and otherwise unremarkable galaxies.

We can see no obvious solution. Adopting some kind of definition based on IR, submm, or radio emission will introduce other problems, mainly to do with ensuring that the measured emission in fact always traces the starburst activity one wishes to define, and the outcome will in many cases not be directly related to what we see in the UV/optical/near-IR regime where traditionally a lot of ``starburst''-oriented work has been performed. Another solution is to restrict the term ``starburst'' to the handful of galaxies which are beyond doubt special systems forming copious amounts of stars (such as UGC~5786 which scores highly in all parameters we considered). A final, fairly unattractive, alternative is to continue using the term ``starburst'' as a rather meaningless adjective to denote a wide-ranging collection of objects (galaxies, regions within galaxies, H{\sc ii} regions, etc.) which have little more in common than that they form massive stars at some rate which by some measure can be called enhanced with respect to something else. Apart from leading to inflation in the use of the definition, this makes it almost impossible to use the term to denote that important class of objects which help the Universe evolve by injecting significant amounts of mechanical and radiation energy and matter into the ISM and IGM over relatively short timescales. Its use should in such cases be limited to the clearly defined scientific question that one is attempting to study, for instance, using enhanced EW to define global starbursts in dwarf galaxies, or using enhanced IR-measured SFR to locate dust-obscured star-forming galaxies in the early Universe.


\section{Conclusions}

We use the SFR and EW measures of a statistically meaningful sample of local galaxies from our \ha GS survey (Paper~I) to study to what extent the presence of a close companion affects the massive SF properties of galaxies. To do this, we first define which galaxies can be considered to have a close companion, and then refine our criteria to select the few galaxies which are interacting or merging. We compare the SFR and EW values of those galaxies with a close companion to those of galaxies without companion. In the second part of the paper, we consider which galaxies from our sample have the highest values for SFR parameters such as SFR, EW, or SFR normalized by area, or the lowest values for gas depletion time. We use this information as input data for a critical discussion of how one can define a ``starburst''. Our main conclusions can be summarized as follows:

\begin{itemize}

\item We find that the presence of a close companion significantly raises the SFR, while only very slightly increasing the EW of the H$\alpha$ emission. This means that although statistically galaxies with close companions form stars at a higher rate, they do this over extended periods of time. A number of tests confirms that this result is robust and not due to sample selection biases.  Across our whole sample, and driven mainly by disk galaxies of intermediate Hubble types, the SFR in galaxies with a close companion is enhanced by a factor of around two compared to those galaxies without companion, whereas the EW enhancement is barely larger than unity. This implies that the SFR increase triggered by the presence of a companion does not occur in the form of an instantaneous ``burst'' of massive SF, but rather as a continuous SF period with a duration of order $10^8$ to $10^9$\,yr.

\item We find a tendency for increased central concentration of the SF as a result of the presence of a close companion, but none of our tests yields a statistically significant result. 

\item We confirm the strong dependence of total and normalized atomic gas mass and of gas depletion timescale on morphological type, but find no relation between the absence or presence of a close companion and the total or normalized \hi\ mass of a galaxy.

\item The fraction of truly interacting or merging galaxies is very small in the local Universe, at around 2\%, and possibly 4\% of bright galaxies. Most of these interacting galaxies have unremarkable star formation properties.

\item A study of the properties of those galaxies in the Survey with the most extreme values for SF indicators such as rate, EW, SFR per area, and gas depletion timescale yields that each of these indicators favors a different subset of galaxies. We use this information to discuss critically the possible definitions of the term starburst to describe galaxies with enhanced SF activity. We conclude that no one starburst definition can be devised which is objective and generally discriminant. Unless one restricts the use of the term ``starburst'' to a very small number of galaxies indeed, the term will continue to be used for a heterogeneous and wide-ranging collection of objects with no physical basis for their classification as starburst.

\end{itemize}


\section*{Acknowledgments}

We acknowledge University of Hertfordshire students Robert Dean and Matthew Chenery for their help with the data collection. We thank Elias Brinks and Gerhardt Meurer for advice on the H\,{\sc i} data, and Isaac Shlosman, Janice Lee, John Beckman, Nils Bergvall, Paola Di Matteo, and Rob Kennicutt for constructive comments on an earlier version of the manuscript. 

This research has made use of the NASA/IPAC Extragalactic Database (NED) which is operated by the Jet Propulsion Laboratory, California Institute of Technology, under contract with the National Aeronautics and Space Administration. We acknowledge the usage of the HyperLeda database (http://leda.univ-lyon1.fr).



\begin{thebibliography}{99}

\bibitem[Alonso-Herrero et al.(2006)]{2006ApJ...650..835A} Alonso-Herrero, 
A., Rieke, G.~H., Rieke, M.~J., Colina, L., P{\'e}rez-Gonz{\'a}lez, P.~G., 
\& Ryder, S.~D.\ 2006, ApJ, 650, 835 

\bibitem[Armus et al.(1987)]{1987AJ.....94..831A} Armus, L., Heckman, T., 
\& Miley, G.\ 1987, AJ, 94, 831 

\bibitem[Arp(1966)]{1966ApJS...14....1A} Arp, H.\ 1966, ApJS, 14, 1 

\bibitem[Arribas et al.(2004)]{2004AJ....127.2522A} Arribas, S., Bushouse, 
H., Lucas, R.~A., Colina, L., \& Borne, K.~D.\ 2004, AJ, 127, 2522

\bibitem[Barton et al.(2000)]{2000ApJ...530..660B} Barton, E.~J., Geller, 
M.~J., \& Kenyon, S.~J.\ 2000, ApJ, 530, 660 

\bibitem[Barton Gillespie et al.(2003)]{2003ApJ...582..668B} Barton 
Gillespie, E., Geller, M.~J., \& Kenyon, S.~J.\ 2003, ApJ, 582, 668 

\bibitem[Berentzen et al.(2004)]{2004MNRAS.347..220B} Berentzen, I., 
Athanassoula, E., Heller, C.~H., \& Fricke, K.~J.\ 2004, \mnras, 347, 220 

\bibitem[Bergvall et 
al.(2003)]{2003A&A...405...31B} Bergvall, N., Laurikainen, E., \& Aalto, S.\ 2003, A\&A, 405, 31

\bibitem[Bigiel et al.(2008)]{2008AJ....136.2846B} Bigiel, F., Leroy, A., 
Walter, F., Brinks, E., de Blok, W.~J.~G., Madore, B., 
\& Thornley, M.~D.\ 2008, AJ, 136, 2846 

\bibitem[Brosch et al.(2004)]{2004MNRAS.349..357B} Brosch, N., Almoznino, 
E., \& Heller, A.~B.\ 2004, MNRAS, 349, 357

\bibitem[Bushouse(1987)]{1987ApJ...320...49B} Bushouse, H.~A.\ 1987, \apj, 
320, 49

\bibitem[]{ButaCat07} Buta, R. J., Corwin, H. G., Odewahn, S. C. 2007, The de
  Vaucouleurs Atlas of Galaxies (Cambridge: Cambridge University
  Press)

\bibitem[Calzetti(2008)]{2008ASPC..390..121C} Calzetti, D.\ 2008, in Proc. Pathways 
Through an Eclectic Universe, Eds. J.~H. Knapen, T.~J. Mahoney, \& A. Vazdekis, ASPC, 390, 121
  
\bibitem[Clements et al.(1996)]{1996MNRAS.279..477C} Clements, D.~L., 
Sutherland, W.~J., McMahon, R.~G., \& Saunders, W.\ 1996, MNRAS, 279, 477 

\bibitem[Cox et al.(2008)]{2008MNRAS.384..386C} Cox, T.~J., Jonsson, P., 
Somerville, R.~S., Primack, J.~R., \& Dekel, A.\ 2008, \mnras, 384, 386 

\bibitem[di Matteo et 
al.(2007)]{2007A&A...468...61D} di Matteo, P., Combes, F., Melchior, A.-L., \& Semelin, B.\ 2007, A\&A, 468, 61 

\bibitem[Di Matteo et al.(2008)]{2008arXiv0809.2592D} Di Matteo, P., 
Bournaud, F., Martig, M., Combes, F., Melchior, A.~-L., 
\& Semelin, B.\ 2008, A\&A, 492, 31

\bibitem[Ellison et al.(2008)]{2008AJ....135.1877E} Ellison, S.~L., Patton, 
D.~R., Simard, L., \& McConnachie, A.~W.\ 2008, AJ, 135, 1877

\bibitem[Gallagher(2005)]{2005ASSL..329...11G} Gallagher, J.~S., III 2005, 
in Starbursts: From 30 Doradus to Lyman Break Galaxies, eds. R. de Grijs \& R.~M. Gonz\'alez Delgado, ASSL, 329, 11 

\bibitem[Garnett(2002)]{2002ApJ...581.1019G} Garnett, D.~R.\ 2002, \apj, 
581, 1019 

\bibitem[Genzel et al.(1998)]{1998ApJ...498..579G} Genzel, R., et al.\ 
1998, \apj, 498, 579 

\bibitem[Gerin et 
al.(1990)]{1990A&A...230...37G} Gerin, M., Combes, F., \& Athanassoula, E.\ 1990, \aap, 230, 37 

\bibitem[Greve et 
al.(2006)]{2006A&A...459..441G} Greve, A., Neininger, N., Sievers, A., \& Tarchi, A.\ 2006, A\&A, 459, 441 

\bibitem[Heckman(2005)]{2005ASSL..329....3H} Heckman, T.~M.\  2005, 
in Starbursts: From 30 Doradus to Lyman Break Galaxies, eds. R. de Grijs \& R.~M. Gonz\'alez Delgado, ASSL, 329, 3

\bibitem[James et 
al.(2004)]{2004A&A...414...23J} James, P.~A., et al.\ 2004, A\&A, 414, 23 (Paper~I)

\bibitem[James et 
al.(2008a)]{2008A&A} James, P.~A., Bretherton, C. F., \& Knapen, J. H., 2008a, submitted to A\&A (Paper~VII)

\bibitem[James et 
al.(2008)]{2008A&A...482..507J} James, P.~A., Knapen, J.~H., Shane, N.~S., Baldry, I.~K., \& de Jong, R.~S.\ 2008b, A\&A, 482, 507 (Paper~IV)

\bibitem[James et 
al.(2008b)]{2008A&A...484..703J} James, P.~A., Prescott, M., \& Baldry, I.~K.\ 2008c, A\&A, 484, 703 

\bibitem[Jogee et al.(2008)]{2008arXiv0802.3901J} Jogee, S., et al.\ 2008, ASPC, 396, 337

\bibitem[Joseph 
\& Wright(1985)]{1985MNRAS.214...87J} Joseph, R.~D., \& Wright, G.~S.\ 1985, MNRAS, 214, 87 

\bibitem[Kapferer et 
al.(2005)]{2005A&A...438...87K} Kapferer, W., Knapp, A., Schindler, S., Kimeswenger, S., \& van Kampen, E.\ 2005, \aap, 438, 87 

\bibitem[Keel(1991)]{1991IAUS..146..243K} Keel, W.~C.\ 1991, in Dynamics of 
Galaxies and Their Molecular Cloud Distributions, eds. F. Combes \& F. Casoli (Kluwer: Dordrecht), p. 243 

\bibitem[Kennicutt 
\& Kent(1983)]{1983AJ.....88.1094K} Kennicutt, R.~C., Jr., \& Kent, S.~M.\ 1983, \aj, 88, 1094 

\bibitem[Kennicutt et al.(1987)]{1987AJ.....93.1011K} Kennicutt, R.~C., 
Jr., Roettiger, K.~A., Keel, W.~C., van der Hulst, J.~M., 
\& Hummel, E.\ 1987, \aj, 93, 1011 

\bibitem[Kennicutt et al.(1994)]{1994ApJ...435...22K} Kennicutt, R.~C., 
Jr., Tamblyn, P., \& Congdon, C.~E.\ 1994, ApJ, 435, 22 

\bibitem[Kennicutt et al.(2005)]{2005ASSL..329..187K} Kennicutt, R.~C., 2005, 
in Starbursts: From 30 Doradus to Lyman Break Galaxies, eds. R. de Grijs \& R.~M. Gonz\'alez Delgado, ASSL, 329, 187

\bibitem[Kennicutt et al.(2008)]{2008ApJS..178..247K} Kennicutt, R.~C., 
Jr., Lee, J.~C., Funes, S.~J., Jos{\'e} G., Sakai, S., 
\& Akiyama, S.\ 2008, \apjs, 178, 247 

\bibitem[Knapen(2005)]{2005A&A...429..141K} Knapen, J.~H.\ 2005, A\&A, 429, 141 

\bibitem[Knapen et 
al.(2006)]{2006A&A...448..489K} Knapen, J.~H., Mazzuca, L.~M., B{\"o}ker, T., Shlosman, I., Colina, L., Combes, F., \& Axon, D.~J.\ 2006, A\&A, 448, 489 

\bibitem[Laine et al.(2002)]{2002ApJ...567...97L} Laine, S., Shlosman, I., 
Knapen, J.~H., \& Peletier, R.~F.\ 2002, ApJ, 567, 97 

\bibitem[Larson 
\& Tinsley(1978)]{1978ApJ...219...46L} Larson, R.~B., \& Tinsley, B.~M.\ 1978, ApJ, 219, 46 

\bibitem[Lee et al.(2007)]{2007ApJ...671L.113L} Lee, J.~C., Kennicutt, 
R.~C., Funes, S.~J., Jos{\'e} G., Sakai, S., 
\& Akiyama, S.\ 2007, ApJ, 671, L113 

\bibitem[Lee et al.(2009)]{2009ApJ...692.1305L} Lee, J.~C., Kennicutt, 
R.~C., Jos{\'e} G.~Funes, S.~J., Sakai, S., 
\& Akiyama, S.\ 2009, ApJ, 692, 1305

\bibitem[Leitherer et al.(1999)]{1999ApJS..123....3L} Leitherer, C., et 
al.\ 1999, ApJS, 123, 3 

\bibitem[Li et al.(2008)]{2008MNRAS.385.1903L} Li, C., Kauffmann, G., 
Heckman, T.~M., Jing, Y.~P., \& White, S.~D.~M.\ 2008, MNRAS, 385, 1903 

\bibitem[Ly et al.(2007)]{2007ApJ...657..738L} Ly, C., et al.\ 2007, ApJ, 
657, 738 

\bibitem[Meurer et al.(2006)]{2006ApJS..165..307M} Meurer, G.~R., et al.\ 
2006, ApJS, 165, 307 

\bibitem[Mihos 
\& Hernquist(1994)]{1994ApJ...425L..13M} Mihos, J.~C., \& Hernquist, L.\ 1994, ApJ, 425, L13 

\bibitem[Mihos 
\& Hernquist(1996)]{1996ApJ...464..641M} Mihos, J.~C., \& Hernquist, L.\ 1996, ApJ, 464, 641 

\bibitem[]{} Nilson, P. 1973, Uppsala general catalogue of galaxies, Uppsala Obs. Ann., vol. 6 (UGC)

\bibitem[\protect\citeauthoryear{Paturel et al.}{2003}]{2003A&A...412...45P} Paturel, G., Petit, C., Prugniel, P., Theureau, G., Rousseau, J., Brouty, M., \& Dubois, P., Cambr{\'e}sy L., 2003, A\&A, 412, 45

\bibitem[Roberts 
\& Haynes(1994)]{1994ARA&A..32..115R} Roberts, M.~S., \& Haynes, M.~P.\ 1994, ARA\&A, 32, 115 

\bibitem[Romano-D{\'{\i}}az et al.(2008)]{2008ApJ...687L..13R} 
Romano-D{\'{\i}}az, E., Shlosman, I., Heller, C., 
\& Hoffman, Y.\ 2008, \apjl, 687, L13 

\bibitem[Rosa-Gonz{\'a}lez et al.(2002)]{2002MNRAS.332..283R} 
Rosa-Gonz{\'a}lez, D., Terlevich, E., 
\& Terlevich, R.\ 2002, MNRAS, 332, 283 

\bibitem[Sanders et al.(1988)]{1988ApJ...325...74S} Sanders, D.~B., Soifer, 
B.~T., Elias, J.~H., Madore, B.~F., Matthews, K., Neugebauer, G., 
\& Scoville, N.~Z.\ 1988, \apj, 325, 74 

\bibitem[Sanders 
\& Mirabel(1996)]{1996ARA&A..34..749S} Sanders, D.~B., \& Mirabel, I.~F.\ 1996, ARA\&A, 34, 749 

\bibitem[Sanders et al.(2003)]{2003AJ....126.1607S} Sanders, D.~B., 
Mazzarella, J.~M., Kim, D.-C., Surace, J.~A., 
\& Soifer, B.~T.\ 2003, AJ, 126, 1607 

\bibitem[Sanders 
\& Ishida(2004)]{2004ASPC..320..230S} Sanders, D., \& Ishida, C.\ 2004, in The Neutral ISM in Starburst Galaxies, eds. S. Aalto, S. H\"uttemeister, \& A. Pedlar, ASPC, 320, 230 

\bibitem[Schmitt(2001)]{2001AJ....122.2243S} Schmitt, H.~R.\ 2001, AJ, 
122, 2243

\bibitem[Scoville et al.(2000)]{2000AJ....119..991S} Scoville, N.~Z., et 
al.\ 2000, AJ, 119, 991 

\bibitem[Shioya et al.(2008)]{2008ApJS..175..128S} Shioya, Y., et al.\ 
2008, ApJS, 175, 128

\bibitem[Smail et al.(2002)]{2002MNRAS.331..495S} Smail, I., Ivison, R.~J., 
Blain, A.~W., \& Kneib, J.-P.\ 2002, MNRAS, 331, 495 

\bibitem[Smith et al.(2007)]{2007AJ....133..791S} Smith, B.~J., Struck, C., 
Hancock, M., Appleton, P.~N., Charmandaris, V., 
\& Reach, W.~T.\ 2007, \aj, 133, 791 

\bibitem[Springel(2000)]{2000MNRAS.312..859S} Springel, V.\ 2000, \mnras, 
312, 859

\bibitem[Struck(1999)]{1999PhR...321....1S} Struck, C.\ 1999, Phys. Rep.,
321, 1

\bibitem[Struck(2007)]{2007IAUS..237..317S} Struck, C.\ 2007, IAU 
Symposium, 237, 317 

\bibitem[van der Hulst et 
al.(1988)]{1988A&A...195...38V} van der Hulst, J.~M., Kennicutt, R.~C., Crane, P.~C., \& Rots, A.~H.\ 1988, A\&A, 195, 38 

\bibitem[Woods 
\& Geller(2007)]{2007AJ....134..527W} Woods, D.~F., \& Geller, M.~J.\ 2007, \aj, 134, 527 

\end{thebibliography}
\end{document}